%% file: manuscript.tex
 \documentclass[final,5p,times,twocolumn,authoryear]{elsarticle}
\nopreprintlinetrue

\usepackage{xcolor}
\usepackage{enumerate}
\usepackage{etoolbox}
\usepackage{mathtools}
\usepackage{amssymb}
\usepackage{amsmath}
\usepackage{multirow}
\usepackage{booktabs}
\usepackage{amssymb}
\usepackage[hidelinks]{hyperref}
\usepackage{flushend}

\usepackage{amsthm}
\newtheorem{lemma}{Lemma}

\theoremstyle{definition}

\newtheorem{remark}{Remark}[section]

\DeclareMathAlphabet\mathbfcal{OMS}{cmsy}{b}{n}
\AtBeginEnvironment{quote}{\par \it}
\newcommand{\rev}[1]{{\leavevmode\color{black}#1}}

\begin{document}
\begin{frontmatter}
\title{Feedback-feedforward Signal Control with Exogenous Demand Estimation\\in Congested Urban Road Networks}
\author{Leonardo~Pedroso\fnref{addressTUe}\corref{correspondingAuthor}} 
\ead{l.pedroso@tue.nl}
\author{Pedro~Batista\fnref{addressISR}}
\author{Markos~Papageorgiou\fnref{addressTUC,addressChina}}
\address[addressTUe]{Control Systems Technology section, Eindhoven University of Technology, The Netherlands}
\address[addressISR]{Institute for Systems and Robotics, Instituto Superior T\'ecnico, Universidade de Lisboa, Portugal}
\address[addressTUC]{Dynamic Systems and Simulation Laboratory, Technical University of Crete, Chania, Greece}
\address[addressChina]{Faculty of Maritime and Transportation, Ningbo University, Ningbo, China}
\cortext[correspondingAuthor]{Corresponding author.}

\begin{abstract}

To cope with uncertain traffic patterns and traffic models, traffic-responsive signal control strategies in the literature are designed to be robust to these uncertainties. These robust strategies still require sensing infrastructure to implement traffic-responsiveness. In this paper, we take a novel perspective and show that it is possible to use the already necessary sensing infrastructure to estimate the uncertain quantities in real time. Specifically, resorting to the store-and-forward model, we design a novel network-wide traffic-responsive strategy that estimates the occupancy and exogenous demand in each link, i.e., entering (exiting) vehicle flows at the origins (destinations) of the network or within links, in real time. Borrowing from optimal control theory, we design an optimal linear quadratic control scheme, consisting of a linear feedback term, of the occupancy of the road links, and a feedforward component, which accounts for the varying exogenous vehicle load on the network. Thereby, the resulting control scheme is a simple feedback-feedforward controller, which is fed with occupancy and exogenous demand estimates, and is suitable for real-time implementation. Numerical simulations for the urban traffic network of Chania, Greece, show that, for realistic surges in the exogenous demand, the proposed solution significantly outperforms tried-and-tested solutions that ignore the exogenous demand.
\end{abstract}


\begin{keyword}
Traffic signal control \sep   Traffic demand estimation \sep  Uncertain traffic demand \sep  Quadratic optimal control \sep Store-and-forward model \sep Feedforward control
\end{keyword}

\end{frontmatter}

\input{./sections/introduction.tex}
\input{./sections/problem_statement.tex}
\input{./sections/estimation.tex}
\input{./sections/control.tex}

\input{./sections/simulation.tex}
\input{./sections/conclusion.tex}
\input{./sections/appendix.tex}
\bibliographystyle{elsarticle-harv} 
\bibliography{references.bib}

\end{document}

%% file: sections/introduction.tex
\section{Introduction}

Oftentimes, infrastructure expansion cannot keep pace with the increasing mobility demand, which leads to traffic congestion even outside rush hour \citep{lomax}. As a result, it is no surprise that significant effort has been put in the design of efficient signal control strategies with the aim of harnessing the existing traffic infrastructure more effectively. Many strategies have been proposed over several decades, which are reviewed in \citet{papageorgiou2003review}. Some of the most popular are SCATS \citep{sims1979scat}, SCOOT \citep{hunt1981scoot}, PRODYN \citep{henry1984prodyn}, UTOPIA \citep{mauro1990utopia}, and RHODES \citep{mirchandani2001real}. Over the last two decades, a host of novel strategies has emerged with emphasis on real-time responsiveness to varying traffic conditions. One of those is the popular and extensively researched Traffic-responsive Urban Control (TUC) strategy. TUC makes use of the store-and-forward model of an urban traffic network, initially presented in \cite{gazis1963oversaturated}, to overcome the exponentially increasing complexity resulting from the discrete nature of the signal control problem. This strategy was initially proposed in \cite{diakaki1999integrated} and has been validated extensively in the field \citep{smaragdis2003application,dinopoulou2005application,KosmatopoulosEtAl2006,kraus2010cost}.  Perimeter feedback-control strategies, which are based on Network Fundamental Diagrams \citep{Daganzo2007,Geroliminis2008}, have also been given significant attention \citep{Keyvan-Ekbatani2012,ChenLiEtAl2022}. In particular, robust perimeter control strategies, which were introduced in \cite{HaddadShraiber2014} and \citet{Haddad2015}, offer robustness to modeling uncertainty. Moreover, some innovative intersection control schemes also account for priority operation of buses \citep{koehler2010simultaneous}. More recently, and with a forward view to vehicle automation, the mixed traffic urban traffic control problem has also been addressed \citep{MaLiEtAl2022,LiYuEtAl2023}, including the joint allocation of crossing for connected automated vehicles, cyclists, and pedestrians \citep{niels2020integrated,niels2020simulation}.


Interestingly, there is still some research effort into fixed-time network-wise signal control strategies \citep{yu2018optimal,zheng2019stochastic,mohebifard2019optimal}. Despite not being traffic responsive, these works employ very complex traffic and routing models. However, traffic patterns may vary dramatically, especially in cities, due to, for instance, a popular event, strike, construction works, or a vehicle collision. Therefore, although fixed-time strategies can cope with recurrent congestion peaks, they may lead to significant performance degradation in those highly dynamical and uncertain scenarios \citep{stevanovic2008stochastic}. As a result, research on traffic-responsive strategies has become more prominent. In \cite{baldi2019simulation}, a traffic-responsive piece-wise linear control law is synthesized making use of a simulation-based procedure. In \cite{zhou2016two}, a promising distributed model-based predictive control scheme is proposed, but it employs a rather simple flow model that does not account for the varying demand on the network. Data-driven neuroevolutionary approaches are also becoming popular \citep{bernas2019neuroevolutionary}, but lack interpretability and require serious processing power. For an extensive review of recent advances in traffic-responsive signal control strategies refer, for instance, to \citet{zhao2011computational} and \citet{qadri2020state}.

It is noted that traffic-responsive strategies are usually designed based on a model which is inevitably uncertain. As a result, some research effort has been dedicated to designing traffic-responsive strategies that are robust to model uncertainties. For example, traffic-responsive signal control strategies robust to uncertainty: (i)~in the demand on the road links are designed in \cite{TettamantiLuspayEtAl2014,ZhouBouyekhfEtAl2015,YeWuEtAl2017}; and (ii)~in model parameters, such as turning rates, are designed in \cite{TettamantiVargaEtAl2011,KomarovskyHaddad2019}. Naturally, since these strategies are traffic-responsive, it is necessary to have access to sensing infrastructure to be able to respond to changes in traffic occupancy. Specifically, all of the aforementioned traffic-responsive strategies rely on real-time feedback of the occupancy in road links, resorting to sensors, such as cameras and inductive loop detectors (see \cite{PadmavathiShanmugapriyaEtAl2010} for a detailed overview of vehicle-detection sensor infrastructure). Since sensing infrastructure is already required to implement a traffic-responsive strategy, one may ask: Is it possible to
\begin{enumerate}[(i)]
		\item use the (already available) sensing infrastructure to \emph{estimate the uncertain quantities} in real time
\end{enumerate}
instead of 
\begin{enumerate}[(i)]
	\setcounter{enumi}{1}
	\item designing strategies that are \emph{robust to uncertainty} in those quantities but unaware of their real-time variation?
\end{enumerate}
This is the research question that this paper focuses on. It is noted that, on the one hand, when designing a robust traffic-responsive signal control strategy, one generally makes assumptions on the bounds or the statistical distribution of the uncertainty. However, although these assumptions may be reasonable during normal operation periods, they do not hold true in cases of strong non-recurrent variation (caused for instance by popular events, strikes, road works, or vehicle collisions), which are not uncommon in an urban environment. On the other hand, an estimation approach can capture such strong variations in the uncertain quantities seamlessly in real time and consider them explicitly in the control decisions.

In this paper, the objective is to leverage a simple, general, transparent, and tried-and-tested traffic-responsive control design procedure, whose synthesis is based on nominal parameters, and employ real-time sensor data to estimate time-varying and highly uncertain traffic patterns to adjust the control action judiciously. In particular, we consider the store-and-forward model, which is used by TUC to synthesize an optimal feedback controller. TUC has been evaluated in practice multiple times in different urban networks \citep{diakaki1999application,dinopoulou2005application,KosmatopoulosEtAl2006} yielding good performance. However, TUC does not account for the time-varying vehicle demand on the road links, which is highly uncertain and dynamical. Our approach is to use loop detector sensing infrastructure to estimate not only the link occupancy but also the net exogenous demand on every road link, i.e., entering (exiting) vehicle flows at the origins (destinations) of the network or within links. To this end, we make use of optimal filtering theory to devise a Kalman filter to estimate those quantities. Moreover, employing tools from optimal control theory, we design a linear quadratic regulator whereby the net exogenous demand is accounted for in a feedforward component. The resulting control scheme is a simple feedback-feedforward controller, which is fed with occupancy and exogenous demand estimates, and is suitable for real-time implementation.

\subsection{Statement of Contributions}

The contributions of this paper are threefold. First, we devise an estimation approach to jointly estimate the occupancy and net exogenous demand on each link, resorting to a single loop detector in each link. Second, we propose a feedback-feedforward controller that makes use of the occupancy and exogenous demand estimates. Third, we show that the feedforward component may lead to significant performance improvements in realistic urban demand scenarios in comparison with state-of-the-art methods that disregard it.

\subsection{Organization}
The remainder of this paper is organized as follows. In Section~\ref{sec:problem_statement}, the store-and-forward model is briefly presented and the estimation and control framework are detailed. In Sections~\ref{sec:estimation} and \ref{sec:control}, estimation and feedback-feedforward control solutions, respectively, are proposed. In Section~\ref{sec:simulation}, the proposed method is validated resorting to numerical simulations of an urban traffic network, and its performance is compared with an analogous signal control method that does not feature a feedforward component. Finally, in Section~\ref{sec:conclusion}, we outline the major conclusions of this work.

\subsection{Notation}
Throughout this paper, the identity and null matrices, both of appropriate dimensions, are denoted by $\mathbf{I}$ and $\mathbf{0}$, respectively. Alternatively, $\mathbf{I}_n$ and $\mathbf{0}_{n\times m}$ are also used to represent the $n\times n$ identity matrix and the $n\times m$ null matrix, respectively. The vector $\mathbf{l_i}$ denotes a column vector whose entries are all set to zero except for the $i$-th one, which is set to $1$. The $i$-th component of a vector $\mathbf{v}\in\mathbb{R}^n$ is denoted by $[\mathbf{v}]_i$, and the entry $(i,j)$ of a matrix $\mathbf{A}$ is denoted by $[\mathbf{A}]_{ij}$. The column-wise concatenation of vectors $\mathbf{x_1},\ldots,\mathbf{x_N}$ is denoted by $\mathrm{col}(\mathbf{x_1},\ldots,\mathbf{x_N})$. The block diagonal matrix whose diagonal blocks are given by matrices $\mathbf{A_1}, ..., \mathbf{A_N}$ is denoted by $\mathrm{diag}(\mathbf{A_1},...,\mathbf{A_N})$. Moreover, $\mathrm{diag}(\mathbf{v})\in \mathbb{R}^{n\times n}$, where $\mathbf{v}\in \mathbb{R}^n$ is a vector, denotes the diagonal matrix whose diagonal entries correspond to the entries of $\mathbf{v}$. Given a symmetric matrix $\mathbf{P}$, $\mathbf{P}\succ\mathbf{0}$  and $\mathbf{P}\succeq\mathbf{0}$ are used to point out that $\mathbf{P}$ is positive definite and positive semidefinite, respectively. 

%% file: sections/problem_statement.tex
\section{Problem statement}\label{sec:problem_statement}
The control and estimation techniques employed in this paper are based on the store-and-forward model of an urban traffic network. In this section, its fundamentals are briefly presented for the sake of completeness. Moreover, we put forward the estimation and control framework that is employed as well as the control objectives. 

\subsection{Store-and-forward model}

The brief presentation of the store-and-forward model follows \citet{diakaki1999integrated,aboudolas2009store}; and \citet{PedrosoBatista2021SignalControl} closely. The topology of the urban network is represented by a directed graph whose nodes correspond to the junctions and whose edges are the directional road links. Such directed graph is denoted by $\mathcal{G} = (\mathcal{V}_\mathcal{G}, \mathcal{E}_\mathcal{G})$, where $\mathcal{V}_\mathcal{G}$ is the set of nodes and $\mathcal{E}_\mathcal{G}$ is the set of edges. Consider a traffic network with $Z$ links and $J$ junctions. Each junction $j \in \{1,\ldots,J\}$ is associated to a vertex $j \in \mathcal{V}_\mathcal{G}$ and a directional road link $z\in \{1,\ldots,Z\}$ directed from junction $i$ to junction $j$ is represented by an edge $e_z = (i,j) \in  \mathcal{E}_\mathcal{G}$, i.e., an edge directed from node $i$ to node $j$ in $\mathcal{G}$. Road links from outside the network towards a junction $j$ (origin links) are represented by an edge $e = (0,j)$. Road links that are directed from a junction towards outside of the network are not considered in the graph representation because their flow cannot be controlled by any junction.

\begin{figure}[t]
	\centering
	\includegraphics[width = \linewidth]{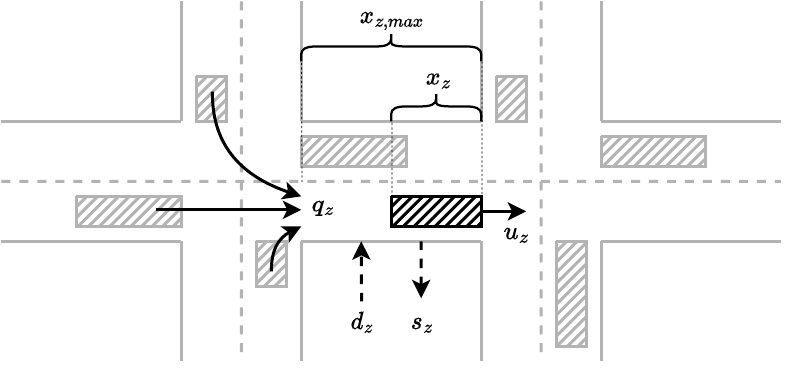}
	\caption{Scheme of the store-and-forward occupancy dynamics of link $z$.}
	\label{fig:saf_scheme}
\end{figure}

Consider a link $z \in \{1,\ldots,Z \}$ and a control discretization period $C$. Denote the number of vehicles in $z$, i.e., its occupancy, at $t = kC$ by  $x_z(k)$. The store-and-forward model describes the occupancy dynamics of link $z$ based on the conservation equation as follows
\begin{equation}\label{eq:flow_conservation}
	x_z(k+1) = x_z(k) + C(q_z(k)-u_z(k)) + C(d_z(k) - s_z(k)),
\end{equation}
where $u_z(k)$ is the outflow of the link $z$ controlled by the traffic signals at the junction downstream; $q_z(k)$ is the inflow to $z$ from other links of the network that enter through the junction upstream; and $d_z(k)$ and $s_z(k)$ are the inflow and outflow, respectively, between link $z$ and infrastructure or unmodeled road unsignalized links that enter and exit, respectively,  link $z$ directly without going through any junction. The maximum admissible number of vehicles in link $z$ is denoted by $x_{z,\mathrm{max}}$.  Fig.~\ref{fig:saf_scheme} depicts a scheme of the store-and-forward occupancy dynamics of link $z$.


Each link $z$ is also characterized by: (i)~a saturation (capacity) flow,  $S_z\in \mathbb{R}_{>0}$; (ii)~turning rates to other links $w\in \{1,\ldots,Z\}$ in the network, $t_{z,w}$; and (iii)~the nominal link exit rate $t_{z,0} \in [0,1[$, which is primarily used for controller synthesis purposes and yields a nominal exit flow 
\begin{equation}\label{eq:s_nom}
	s^{\text{nom}}_z(k)=t_{z, 0} q_z(k).
\end{equation}
Define, for simplicity, the turning rate matrix $\mathbf{T}\in \mathbb{R}^{Z\times Z}$, whose entries are defined as $[\mathbf{T}]_{zw} := t_{w,z}$ with $z,w\in \{1,\ldots,Z\}$, and the exit rate vector $\mathbf{t_0} := [t_{0,1} \; \cdots\; t_{0,Z}]^\top$. At any instant, the traffic network model is characterized by the triplet $(\mathcal{G},\mathbf{T}, \mathbf{t_0})$, which may be time-varying. For the sake of simplicity, we consider, henceforth, that it is time-invariant. 


Even though the nominal exit rates $t_{z,0}$ offer a nominal model for $s_z(k)$ in \eqref{eq:s_nom} that is useful for control synthesis purposes, the exogenous load on the network may cause strong variations about the nominal flows. As a result, we define the net unmodeled exogenous demand on link $z$ at time $t = kC$ as
\begin{equation}\label{eq:def_exogenous_demand}
e_z(k) := d_z(k)-(s_z(k)-s^{\text{nom}}_z(k)).
\end{equation}
Note that $e_z(k)$ may be negative if, due to some external factor, a higher than nominal vehicle flow exits link $z$ towards unmodeled infrastructure.
Substituting \eqref{eq:def_exogenous_demand} in \eqref{eq:flow_conservation} yields
\begin{equation}
x_z(k+1) = x_z(k) + C((1-t_{z,0})q_z(k)-u_z(k)) + Ce_z(k).
\end{equation} 

\begin{remark}
	Instead of defining the exogenous demand as in \eqref{eq:def_exogenous_demand}, one could disregard the nominal exit flow model \eqref{eq:s_nom} and define the exogenous flow on link $z$ as $d_z(k)-s_z(k)$ (this is equivalent to setting $t_{z,0} = 0$). We did not opt for this approach for two reasons. First, notice that disregarding the nominal model for the exit flow in  \eqref{eq:s_nom} results in a loss of information that may be used in a propagation step to obtain better estimation performance. Second, control-oriented store-and-forward models generally account for \eqref{eq:s_nom}, e.g., \cite{aboudolas2009store}, thus \eqref{eq:def_exogenous_demand} arises seamlessly as a disturbance in such models. This aspect will become evident in Section~\ref{sec:control}.
\end{remark}

The signal control operation at each junction is based on cycles of duration $C$. In each cycle, each junction $j$ has a number of predefined stages, each denoted by a unique integer $s\in \mathcal{F}_j \subset \{1,\ldots,S\}$, whereby each stage gives right of way (r.o.w.) to a set of links directed towards that junction. We define the stage matrix $\mathbf{S}\in \mathbb{R}^{Z\times S}$ as
\begin{equation}
	[\mathbf{S}]_{zs} := \begin{cases}
		1, & \quad \text{if link }$z$\text{ has r.o.w. at stage } $s$ \\
		0, & \quad \text{otherwise}.
	\end{cases}
\end{equation}
During each cycle $k$, each stage $s$ has an associated green-time $g_s(k)$, which is the control variable available to the network operator to regulate the flow of vehicles. The green-time of each stage $s$ must be no-smaller than the minimum permissible green-time $g_{s,\mathrm{min}} \in \mathbb{R}_{\geq0}$, i.e.,
\begin{equation}\label{eq:constr_gs_1}
	g_s(k) \geq g_{s,\mathrm{min}}, \quad \forall k\in \mathbb{N}_0 \; \forall s\in \{1,\ldots,S\}
\end{equation}
and 
\begin{equation}\label{eq:constr_gs_2}
	\sum_{s\in \mathcal{F}_j} g_s(k) +L_j = C, \quad   \forall k\in \mathbb{N}_0 \; \forall j\in \{1,\ldots,J\},
\end{equation}
where $L_j$ is the total duration of all inter-green times of junction $j$ that are introduced for safety reasons.

The prime simplification introduced by the store-and-forward model is modeling the green-red switching of the traffic lights within a control cycle as a continuous flow of vehicles. Taking into account the switching behavior, if the link has r.o.w., the outflow of vehicles of each link $z$ equals the saturation flow $S_z$, and, otherwise, the outflow is null.  Since our sample time equals $C$, the cycle time, the outflow for the whole cycle equals the average flow, i.e., 
\begin{equation}\label{eq:LTI_saf_u}
	u_z(k) = S_z\sum_{s:[\mathbf{S}]_{zs}\neq 0} g_s(k)/C, \quad z\in \{1,\ldots,Z\}.
\end{equation} 
Note, however, that \eqref{eq:LTI_saf_u} does not account for the following two cases, in which the flow $u_z(k)$ cannot materialize: (i) There are not enough vehicles in link $z$; or (ii) A downstream link is blocked, hence vehicles cannot cross the junction even though they have r.o.w.

Therefore, under this model, the occupancy dynamics can be written for the whole network as the linear time-invariant (LTI) system
\begin{equation}\label{eq:LTI_saf}
\mathbf{x}(k+1) = \mathbf{x}(k) + \mathbf{B_g}\mathbf{g}(k) + C\mathbf{e}(k), 
\end{equation}
where  $\mathbf{x}(k)\!:=\!\operatorname{col}(x_1(k), \ldots, x_Z(k))\! \in\! \mathbb{R}^Z$,  $\mathbf{g}(k)\!:=\!\operatorname{col}(g_1(k), \ldots, g_S(k)) \!\in\! \mathbb{R}^S$, $\mathbf{e}(k):=\operatorname{col}(e_1(k), \ldots, e_Z(k)) \in \mathbb{R}^Z$, and
$\mathbf{B}_{\mathbf{g}}=((\mathbf{I}_Z-\operatorname{diag}(\mathbf{t}_{\mathbf{0}})) \mathbf{T}-\mathbf{I}_Z) \operatorname{diag}(S_1, \ldots, S_Z) \mathbf{S} \in \mathbb{R}^{Z \times S}$. For more details on the derivation of this system, refer to \citet[Appendix A]{PedrosoBatista2021SignalControl}. Furthermore, for an illustrative example of modeling a network employing the store-and-forward model, refer to \citet[Section~2.3]{PedrosoBatista2021SignalControl}.

\subsection{Nonlinear model}

Even though the store-and-forward LTI model is very simple and useful for control synthesis purposes, it does not enforce the hard occupancy constraints in the links, i.e., $0 \leq x_z(k) \leq x_{z, \max}, \forall z \in \{1,\ldots,Z\}$ due to \eqref{eq:LTI_saf_u}. Thus, for simulation purposes, the occupancy dynamics are better approximated by a more elaborate discrete-time nonlinear model adapted from the store-and-forward modeling approach. We employ a model inspired by the one proposed in \cite{aboudolas2009store}. Consider a simulation sampling time $T \ll C$, assuming for the sake of simplicity that $C/T \in \mathbb{N}$.  On one hand, to satisfy the nonnegativeness of the occupancy of each link, the outflow of vehicles in link $z$ at each interval of duration $T$ is limited by $x_z(k_T)/T$, where, by a slight abuse of notation, $k_T$ is the discrete-time instant at $t = k_TT$ employed to describe $x_z$ with a sampling time $T$. On the other hand, to satisfy the link's limited space capacity constraints, back-holding of flows feeding a saturated link is modeled, which is conform traffic rules. Indeed, whenever a link downstream of a certain link $z$ is near full occupancy, then the outflow of link $z$ has to be null even if it has r.o.w., since the vehicles have no space to proceed downstream. Based on these considerations, the model 
\begin{equation}\label{eq:LTI_sim_nl}
	\mathbf{x}(k_T+1) = \mathbf{x}(k_T)+(T/C)\mathbf{B_u}\mathbf{u_{nl}}(k_T) + T\mathbf{e_{nl}}(k_T)
\end{equation} 
is adopted, where $\mathbf{B}_{\mathbf{u}}=C((\mathbf{I}_Z-\operatorname{diag}\left(\mathbf{t}_{\mathbf{0}})) \mathbf{T}-\mathbf{I}_Z\right)$
and  $\mathbf{u_{\text{nl}}}(k_T) := \mathrm{col}(u_{\text{nl},1}(k_T),...,u_{\text{nl},Z}(k_T))\in \mathbb{R}^{Z}$ with elements
\begin{equation}\label{eq:u_nl}
	u_{\text{nl},z}(k_T) = \begin{cases}
		0 \:,\; \quad \quad \quad \exists w: t_{w,z} \neq 0 \land x_w(k_T)>c_{\text{ug}}x_{w,\text{max}}\\
		\mathrm{min}\left\{\frac{x_z(k_T)}{T},  u_z\left(k = \left \lfloor \frac{k_TT}{C} \right\rfloor\right) \right\}, \quad\quad \quad \text{otherwise},
	\end{cases}
\end{equation}
where $u_z(k)$ is the command action updated every cycle $C$ given the stage green times that follow from \eqref{eq:LTI_saf_u}. Notice that \eqref{eq:u_nl} incorporates the mentioned occupancy constraints with $c_{\text{ug}} \in ]0,1[$ being a parameter to be tuned in order to adjust the sensitivity of upstream back-holding. The extension in relation to the model proposed in \cite{aboudolas2009store} is portrayed in the term $\mathbf{e_{\text{nl}}}(k_T) := \mathrm{col}(e_{\text{nl},1}(k_T),...,e_{\text{nl},Z}(k_T))\in \mathbb{R}^{Z}$, which is the factual link demand that ensures that the exogenous demand does not lead to violation of the occupancy constraints. To this end, it is necessary to keep record of the number of exogenous-demand vehicles that were blocked from entering the link due to the maximum occupancy constraint. To this end, denote the difference between the requested vehicle demand and the available space capacity at time instant $k_T$ as
\begin{equation}
	\!\Delta_z(k_T) := e_z(k_T)T - \left(x_{z,\mathrm{max}}\!\!- \!x_z(k_T)- (T/C)\mathbf{l_z}\!^\top\mathbf{B_u}\mathbf{u_{nl}}(k_T) \right)\!.
\end{equation}
If $\Delta_z(k_T)$ is positive, then the exogenous vehicle demand on the link would exceed the space capacity, hence it cannot enter the link. Otherwise, the demand on the link, along with previously blocked vehicles, may enter the link. We denote the accumulated number of blocked vehicles in link $z$ at time instant $k_T$ by $x_{\text{b},z}(k_T)$, whose dynamics are given by a conservation equation
\begin{equation}
	x_{\text{b},z}(k_T+1) = x_{\text{b},z}(k_T) - T\left(e_{\text{nl},z}(k_T)-e_z(k_T)\right),
\end{equation}
where we have the factual demand according to the above considerations as follows
\begin{equation}
		e_{\text{nl},z}(k_T) = \begin{cases}
				e_z(k_T)- \Delta_z(k_T)/T,   \quad & \Delta_z(k_T) \geq 0 \\
			e_z(k_T)+  	\mathrm{min}\left\{-\Delta_z(k_T),x_{\text{b},z}(k_T)\right\}\!/T, \quad & \Delta_z(k_T) < 0.
	\end{cases}
\end{equation}


\subsection{Objectives and framework}

The goal of this work is to synthesize a feedback-feedforward control approach based on the store-and-forward model \eqref{eq:LTI_saf}. Recall that $\mathbf{e}(k)$ in \eqref{eq:LTI_saf} accounts for the net unmodeled exogenous inflow and outflow between infrastructure and the link, i.e., for flows that appear in the modeled network without crossing any modeled junction. As a matter of fact, this demand can vary strongly and may not be easily predictable. Consider, for example, that there is a football match or a concert in the city center. In this scenario, a few hours before the event,  there will be an extraordinarily high positive exogenous demand in the outer links and a negative high exogenous demand in the links close to the event that represents the flow of vehicles to parking infrastructure. Conversely, after the event, a high exogenous demand in the links close to the event is to be expected.

In \cite{diakaki1999integrated}, the TUC method was first proposed. The main idea behind it is to design an optimal feedback controller for the LTI system \eqref{eq:LTI_saf}, whereby a constant feedforward component, labeled historic green time, is employed to account for an average exogenous demand. The traffic-responsive properties of TUC, endowed by the feedback loop, allow for good disturbance rejection and lead to good performance overall. Our objective is to investigate whether estimating and actively accounting for the exogenous disturbance in the control design  via a feedforward term unlocks significant performance.

The extension of TUC with a feedforward term, that is proposed in this paper, is labeled TUC-FF. The derivation of TUC-FF follows two main stages. First, in Section~\ref{sec:estimation}, we devise a filter to estimate both the occupancy (to be used in the feedback portion of the controller as usual) and net unmodeled exogenous demand  $e_z(k)$ (to be used in the novel feedforward term) on every link $z$. Taking into account the economic limitations of installing sensing infrastructure, we assume the existence of a single loop detector in every modeled link in the network. Second, in Section~\ref{sec:control}, we derive a feedback-feedforward control solution that is fed with the occupancy and exogenous demand estimates.


The performance of the control methods is evaluated making use of three metrics: (i)~the total time spent (TTS) including the blocked demand due to full links downstream
\begin{equation}
	\operatorname{TTS} =T \sum_{k_T} \sum_{z=1}^Z \left(x_z(k_T) + x_{\text{b},z}(k_T)\right),
\end{equation}
(ii)~the relative queue balance (RQB)
\begin{equation}
\mathrm{RQB}=\sum_{k_T} \sum_{z=1}^Z \frac{x_z^2(k_T)}{x_{z, \max }},
\end{equation}
and (iii)~the total time blocked (TTB)
\begin{equation}
		\operatorname{TTB}= T \sum_{k_T} \sum_{z=1}^Z x_{\text{b},z}(k_T).
\end{equation}
The RBQ was first proposed in \cite{aboudolas2009store}. 

%% file: sections/estimation.tex
\section{Estimation of occupancy and exogenous demand}\label{sec:estimation}

In this section, we devise an approach to estimate the occupancy and net exogenous demand of every link. Due to the high cost and complexity of setting up sensing infrastructure, we consider that only a single loop detector in the middle of every road link is available. As studied thoroughly in \cite{vigos2008real} and \citet{vigos2010simplified}, 
the local time-occupancy measurement that is collected by a single loop detector in the middle of a road link can be translated into a low-biased measurement of link space occupancy. Indeed, therein it is shown that the expected occupancy of a whole link can be given as the space average of the local time-averaged occupancy at each specific location along the link. The time-averaged occupancy at a given location can be known if a loop-detector is available at that specific location.  On signalized links, the average speed of the vehicles tends to monotonically decrease as the vehicles move along the link towards the queue. Because of that, crucially, it is shown that, under realistic settings, the local time-averaged occupancy in the middle of the link is a low-biased estimate of the occupancy of the link. Consider a sampling time $E \gg T$ and assume for the sake of simplicity that  $C/E \in \mathbb{N}$ and $E/T \in \mathbb{N}$. For filter design purposes, we assume that the measurement noise is Gaussian, uncorrelated, and independent among the sensors. Thus, the occupancy measurement on link $z$ is denoted by $y_z(k_E)$ and modeled as
\begin{equation}\label{eq:sensor_output}
	y_z(k_E) = x_z(k_E) + v_z(k_E),
\end{equation}
where $k_E$ is the discrete time instant at $t = k_EE$ and $v_z(k_E)$ is a zero-mean white Gaussian process whose covariance is given by $R_z \in \mathbb{R}_{>0}$, see \cite{vigos2008real} and \citet{vigos2010simplified} for more details. It is worth remarking that, in practice, $v_z(k_E)$ may not be exactly Gaussian and be slightly correlated in time. However, the Kalman filter is well-known to yield good estimation performance \rev{when these effects are small. Otherwise, different techniques can be used (e.g.,  \cite[Chapter 4.5]{gelb1974applied}).} Indeed, acceptable estimation performance is achieved in the simulation results in Section~\ref{sec:simulation}, where colored noise is considered.

If the average outflows in each time interval $[k_EE,(k_E+1)E[$ on every link of the network $\mathbf{u}(k_E) := \mathrm{col}(u_{1}(k_E),...,u_{Z}(k_E))\in \mathbb{R}^{Z}$ were known, then one could easily write the estimation dynamics for a sampling time $E$. If a loop detector were available at the end of every link, then $\mathbf{u}(k_E)$ could be measured directly. However, in our framework, that is not available. As a result, we introduce the following approximation, where we assume that the average outflow in a link $z$ in each time interval $[k_EE,(k_E+1)E[$ is given by 
\begin{equation}\label{eq:estimated_u}
	u_{\mathrm{est},z}(k_E) = \begin{cases}
		0 \:,\; \quad \quad \quad \quad \; \;\exists w: t_{w,z} \neq 0 \land x_w(k_T)>c_{ug}x_{w,\text{max}}\\
		\mathrm{min}\left\{\frac{x_z(k_E)}{E}, \!\!\!  \sum\limits_{s:[\mathbf{S}]_{zs}\neq 0} \!\!\!g_s\left(\left(k = \left \lfloor \frac{k_EE}{C} \right\rfloor\right)\right) \frac{S_z}{C} \right\}\!,\;\; \text{otherwise},
	\end{cases}
\end{equation}
which resembles the nonlinear model \eqref{eq:u_nl}, differing only in the sampling rate. However, given the high sampling period $E$, this is going to introduce process noise in the estimation dynamics, i.e., 
\begin{equation}\label{eq:est_dyn}
	x_z(k_E+1) = x_z(k_E) + (E/C)\mathbf{l}_z^\top\mathbf{B_u}\mathbf{u_{\text{est}}}(k_E) + Ee_z(k_E) + w_{\mathrm{x},z}(k_E), 
\end{equation}
where $\mathbf{u}_{\mathrm{est}}(k_E) := \mathrm{col}(u_{\mathrm{est},1}(k_E),...,u_{\mathrm{est},Z}(k_E))\in \mathbb{R}^{Z}$ and the process noise $w_{\mathrm{x},z}(k_E)$ is modeled by a zero-mean white Gaussian process whose covariance is given by $Q_{\mathrm{x},z} \in \mathbb{R}_{>0}$. Moreover, we assume that the net exogenous demand is slowly time-varying and, thus, evolves according to a random walk
\begin{equation}
	e_z(k_E+1) = e_z(k_E) + w_{\mathrm{e,z}}(k_E),
\end{equation}
where $w_{\mathrm{e,z}}(k_E)$ is modeled by a zero-mean white Gaussian process whose covariance is given by $Q_{\mathrm{e},z} \in \mathbb{R}_{>0}$.

A classical Kalman filtering approach may now be followed. Denote the predicted estimate of the occupancy and exogenous demand at discrete time instant $k_E+1$ by $\hat{x}_z(k_E+1|k_E)$ and $\hat{e}_z(k_E+1|k_E)$, respectively, and the filtered estimates by $\hat{x}_z(k_E+1|k_E+1)$ and $\hat{e}_z(k_E+1|k_E+1)$, respectively.  The prediction step can be written as
\begin{equation}
	\begin{bmatrix}
			\hat{x}_z(k_E+1|k_E)\\
			\hat{e}_z(k_E+1|k_E)
	\end{bmatrix} = \underbrace{\begin{bmatrix} 1 & E\\ 0 & 1 	\end{bmatrix}}_{\mathbfcal{A}:=}
\begin{bmatrix}
	\hat{x}_z(k_E|k_E)\\
	\hat{e}_z(k_E|k_E)
\end{bmatrix} + 
\begin{bmatrix}
	(E/C)\mathbf{l_z}\!^\top\mathbf{B_u}\hat{\mathbf{u}}_{\text{est},z}(k_E)\\
	0
\end{bmatrix}
\end{equation}
for $z \in  \{1,\ldots,Z\}$, where $\hat{\mathbf{u}}_{\text{est},z}(k_E)$ is the estimated approximated outflow in the links, which is computed according to \eqref{eq:estimated_u} making use of the filtered estimates $\{\hat{x}_z(k_E|k_E)\}_{z\in \{1,\ldots,Z\}}$, and $\mathbfcal{A}\in \mathbb{R}^{2\times 2}$ is the augmented state transition matrix. Note that, from \eqref{eq:sensor_output}, the output matrix is given by $\mathbf{C} = 1$. Since the pair $(\mathbfcal{A},\mathbf{C})$ is observable, it is straightforward to synthesize the optimal Kalman gain $\mathbf{K}_z := [K_{\mathrm{x},z} \;  K_{\mathrm{e},z} ]^\top$ given an augmented process noise covariance matrix $\mathbfcal{Q}_z := \mathrm{diag}(Q_{\mathrm{x},z},Q_{\mathrm{e},z})$ and sensor noise covariance matrix $\mathbfcal{R}_z = R_z$, for $z \in \{1,\ldots,Z\}$. The filtering step is, then, given by 
\begin{equation}
	\begin{cases}
		\hat{x}_z(k_E\!+\!1|k_E\!+\!1) \!= \!\hat{x}_z(k_E\!+\!1|k_E) +  K_{\mathrm{x},z}(y_z(k_E\!+\!1)-\hat{x}_z(k_E\!+\!1|k_E))\\
		\hat{e}_z(k_E\!+\!1|k_E\!+\!1) \!=\! 	\hat{e}_z(k_E\!+\!1|k_E) + K_{\mathrm{e},z}(y_z(k_E\!+\!1)-\hat{x}_z(k_E\!+\!1|k_E)).
	\end{cases}
\end{equation}
Since the pair $(\mathbfcal{A},\mathbf{C})$ is observable, it is possible to conclude that a single loop detector in each road link is, indeed, enough to estimate simultaneously the occupancy and exogenous demand in each road link of the whole network.

%% file: sections/control.tex
\section{Feedback-feedforward signal control}\label{sec:control}

In this section, an optimal feedback-feedforward controller is derived for the LTI system \eqref{eq:LTI_saf}. For controller synthesis purposes, we assume, henceforth, that the exogenous demand is constant. Afterwards, the feedforward component that stems from the exogenous demand is periodically updated with the time-varying demand estimates that are cascaded from the filtering solution developed in Section~\ref{sec:estimation}. For that reason, instead of \eqref{eq:LTI_saf}, we consider the LTI system 
\begin{equation}\label{eq:LTI_saf_cte}
	\mathbf{x}(k+1) = \mathbf{x}(k) + \mathbf{B_g}\mathbf{g}(k) + C\bar{\mathbf{e}}, 
\end{equation}
where $\bar{\mathbf{e}}$ is the constant net exogenous demand. We employ results of optimal control theory to derive a control law for \eqref{eq:LTI_saf_cte}. 

First, consider the following result in \cite{PedrosoBatista2021SignalControl} on the controllability of \eqref{eq:LTI_saf_cte}.
\begin{lemma}[{\citet[Proposition~3.1]{PedrosoBatista2021SignalControl}}]\label{lemma:controllability}
	Consider a feasible traffic network characterized by $(\mathcal{G},\mathbf{T},\mathbf{t_0})$ and a minimum complete stage strategy characterized by stage matrix $\mathbf{S}$.  Let $\mathbfcal{C}$ be the controllability matrix of the store-and-forward LTI system \eqref{eq:LTI_saf_cte}. Then, $\mathrm{rank}(\mathbfcal{C}) = S\leq Z$.
\end{lemma} 

The definition of a feasible traffic network and a minimum complete stage strategy can be found in \cite{PedrosoBatista2021SignalControl}, which are very mild conditions that physically meaningful urban traffic networks and stage structures abide by. Most importantly, from Lemma~\ref{lemma:controllability}, unless $S = Z$, which is rarely the case, the store-and-forward LTI system \eqref{eq:LTI_saf_cte} is not controllable. To circumvent this issue, the technique proposed in \cite{PedrosoBatista2021SignalControl} is employed. Denote the controllability matrix of system \eqref{eq:LTI_saf_cte} by $\mathbfcal{C}$. Leveraging the Canonical Structure Theorem \cite[Chap. 18]{rugh1996linear}, one may decompose \eqref{eq:LTI_saf_cte} into controllable and uncontrollable components. Consider the linear transformation $\mathbf{z}(k) = \mathbf{W}^{-1}\mathbf{x}(k)$, whereby the first $S$ columns of $\mathbf{W}\in \mathbb{R}^{Z\times Z}$ are a basis of the column space of  $\mathbfcal{C}$ and the last $Z-S$ are such that all columns of $\mathbf{W}$ form a basis of $\mathbb{R}^Z$. It follows that \eqref{eq:LTI_saf_cte} can be decomposed as
\begin{equation}\label{eq:cst_components}
	\begin{cases}
			\mathbf{z_1}(k+1) = \mathbf{z_1}(k) + \mathbf{{B}_{g1}}\mathbf{g}(k) + C\mathbf{\bar{e}_{1}}\\
			\mathbf{z_2}(k+1) = \mathbf{z_2}(k) + C\mathbf{\bar{e}_{2}},
	\end{cases}
\end{equation} 
where $\mathbf{z_1}(k) \in \mathbb{R}^S$ and $\mathbf{z_2}(k) \in \mathbb{R}^{Z-S}$ are the states of the controllable and uncontrollable components, respectively; $\mathbf{W}^{-1}\mathbf{B_g} = \left[\mathbf{B_{g1}}^\top \; \mathbf{0}^\top_{(Z-S)\times S}\right]^\top$; and $\mathbf{\bar{e}_{1}} \in \mathbb{R}^{S}$ and $\mathbf{\bar{e}_{2}} \in \mathbb{R}^{Z-S}$ are given by $\mathbf{W}^{-1}\mathbf{\bar{e}} = \left[\mathbf{\bar{e}_1}^\top \;\; \mathbf{\bar{e}_2}^\top\right]^\top$. Although the uncontrollable component in \eqref{eq:cst_components} can seemingly grow unbounded, the link capacity constraints ensure that the occupancy of the links remains bounded. Thus, one can only aim to synthesize a feedback-feedforward controller for the controllable component of \eqref{eq:cst_components}.

Consider a cost function
\begin{equation}\label{eq:cost}
	\begin{split}
		J &= \! \sum_{k =0}^{\infty} \left( \mathbf{z_1}^\top(k)\mathbf{Q_1}\mathbf{z_1}(k) + \mathbf{g}^\top(k)\mathbf{R} \mathbf{g}(k) \right)\\
		&= \!\sum_{k = 0}^{\infty} \left( \mathbf{x}(k)^\top\mathbf{W}^{-\top}\!\begin{bmatrix}\mathbf{Q_1} & \mathbf{0}_{S\times (Z-S)} \\ \mathbf{0}_{(Z-S)\times S} \!\!& \!\!\mathbf{0}_{(Z-S)\times (Z-S)}\end{bmatrix}\mathbf{W}^{-1}\mathbf{x}(k) + ||\mathbf{g}(k)||_\mathbf{R}  \right),
	\end{split}		
\end{equation}
where $\mathbf{Q_1} \succeq \mathbf{0}$ and $\mathbf{R} \succ \mathbf{0}$ are selected matrices of appropriate dimensions. As suggested in \cite{aboudolas2009store}, the occupancy weighting matrix should be set to $\mathrm{diag}(1/x_{1,\text{max}},\ldots,1/x_{Z,\text{max}})$ as a means of minimizing relative occupancy of the links. That is analogous to weighting the controllable component with
\begin{equation}
	\mathbf{Q_1} = \begin{bmatrix}\mathbf{I}_S \!\!&\!\! \mathbf{0}_{S\times (Z-S)}\end{bmatrix}\mathbf{W}^\top\mathrm{diag}(1\!/x_{1,\text{max}},\ldots,1\!/x_{Z,\text{max}})\mathbf{W} \begin{bmatrix}\mathbf{I}_S \\ \mathbf{0}_{(Z-S)\times S\!}\end{bmatrix}\!.
\end{equation}
For more details on the decomposition technique, refer to \citet[Section~3]{PedrosoBatista2021SignalControl}.

Now, one can make use of the following result to synthesize a feedback-feedforward control law for the controllable component of \eqref{eq:LTI_saf_cte}.

\begin{lemma}\label{lemma:fb-ff}
	Consider the controllable component  of \eqref{eq:LTI_saf_cte} in \eqref{eq:cst_components} and the infinite-horizon cost function in \eqref{eq:cost}. The optimal control law is given by
	\begin{equation}\label{eq:fb-ff}
		\mathbf{g}(k) = - \mathbf{K_1}\mathbf{z_1}(k) - C\mathbf{{K}_{e1}}\mathbf{\bar{e}_1},
	\end{equation}
	where $ \mathbf{K}_1$ follows from the solution to the discrete algebraic Ricatti equation (DARE)
	\begin{equation}
		\begin{cases}
			\mathbf{P} = \mathbf{Q_1} + \mathbf{P}(\mathbf{I}-\mathbf{B_{g1}}\mathbf{K})\\
			\mathbf{K_1} = (\mathbf{R}+\mathbf{B_{g1}}^\top \mathbf{P}\mathbf{B_{g1}})^{-1}\mathbf{B_{g1}}^\top\mathbf{P}
		\end{cases}
	\end{equation} 
and 
\begin{equation}
	\mathbf{{K}_{e1}} = (\mathbf{R}+\mathbf{B_{g1}}^\top \mathbf{P}\mathbf{B_{g1}})^{-1}\mathbf{B_{g1}}^\top\left(\mathbf{I}-(\mathbf{I}-\mathbf{B_{g1}}\mathbf{K_1})^\top\right)^{-1}\mathbf{P}.
\end{equation}
\end{lemma}
\begin{proof}
	See \ref{app:proof_fb-ff}.
\end{proof}

Finally, from Lemma~\ref{lemma:fb-ff}, one may cascade the occupancy and exogenous demand estimates into the control law \eqref{eq:fb-ff} to yield
\begin{equation}\label{eq:ctrl_law}
		\mathbf{g}(k) = - \mathbf{K}\left[\hat{\mathbf{x}}(k)\right]_{\mathbf{0}}^{\mathbf{x}_{\text{max}}} - C\mathbf{{K}_{e}}\hat{\mathbf{e}}(k),
\end{equation}
where $\mathbf{K}=\mathbf{K}_{\mathbf{1}}\left[ \mathbf{I}_S \; \mathbf{0}_{S \times(Z-S)} \right]\!\mathbf{W}^{-1}$,  $\mathbf{K_e}=\mathbf{K_{e1}}\left[ \mathbf{I}_S \; \mathbf{0}_{S \times(Z-S)} \right]\!\mathbf{W}^{-1}$, $\hat{\mathbf{e}}(k):=\operatorname{col}(\hat{e}_1(k), \ldots, \hat{e}_Z(k)) \in \mathbb{R}^Z$,  and $\left[\hat{\mathbf{x}}(k)\right]_{\mathbf{0}}^{\mathbf{x}_{\text{max}}}$ denotes the vector of occupancy estimates $\mathbf{x}(k)$ whereby every component of index $z\in \{1,\ldots,Z\}$ is saturated between $0$ and $x_{z,\text{max}}$.

However, we are yet to incorporate the hard constraints \eqref{eq:constr_gs_1} and \eqref{eq:constr_gs_2} on the green-times. As proposed in \cite{diakaki1999integrated} and \cite{aboudolas2009store}, one may adjust the control action from the control law \eqref{eq:ctrl_law} such that they are satisfied. That amounts to solving 
\begin{equation}\label{eq:knapsackLQ_2}
	\begin{aligned}
		& \underset{\tilde{g}_s(k),\: s\in \mathcal{F}_j}{\text{minimize}}
		& & \frac{1}{2}\sum_{s\in F_f} \left(\tilde{g}_s(k)-g_s(k)\right)^2\\
		& \text{subject to}
		& & \tilde{g}_s(k) \geq g_{s,min}  \:, s\in \mathcal{F}_j\\
		& & & \sum_{s\in \mathcal{F}_j} \tilde{g}_s(k) + L_j = C,
	\end{aligned}
\end{equation}
for $j\in \{1,\ldots,J\}$. Since \eqref{eq:knapsackLQ_2} is a quadratic continuous knapsack problem, its solution can be determined making use of very efficient algorithms \citep{helgason1980polynomially}.

%% file: sections/simulation.tex
\section{Simulation results}\label{sec:simulation}

\begin{figure}[t]
	\centering
	\includegraphics[width =\linewidth]{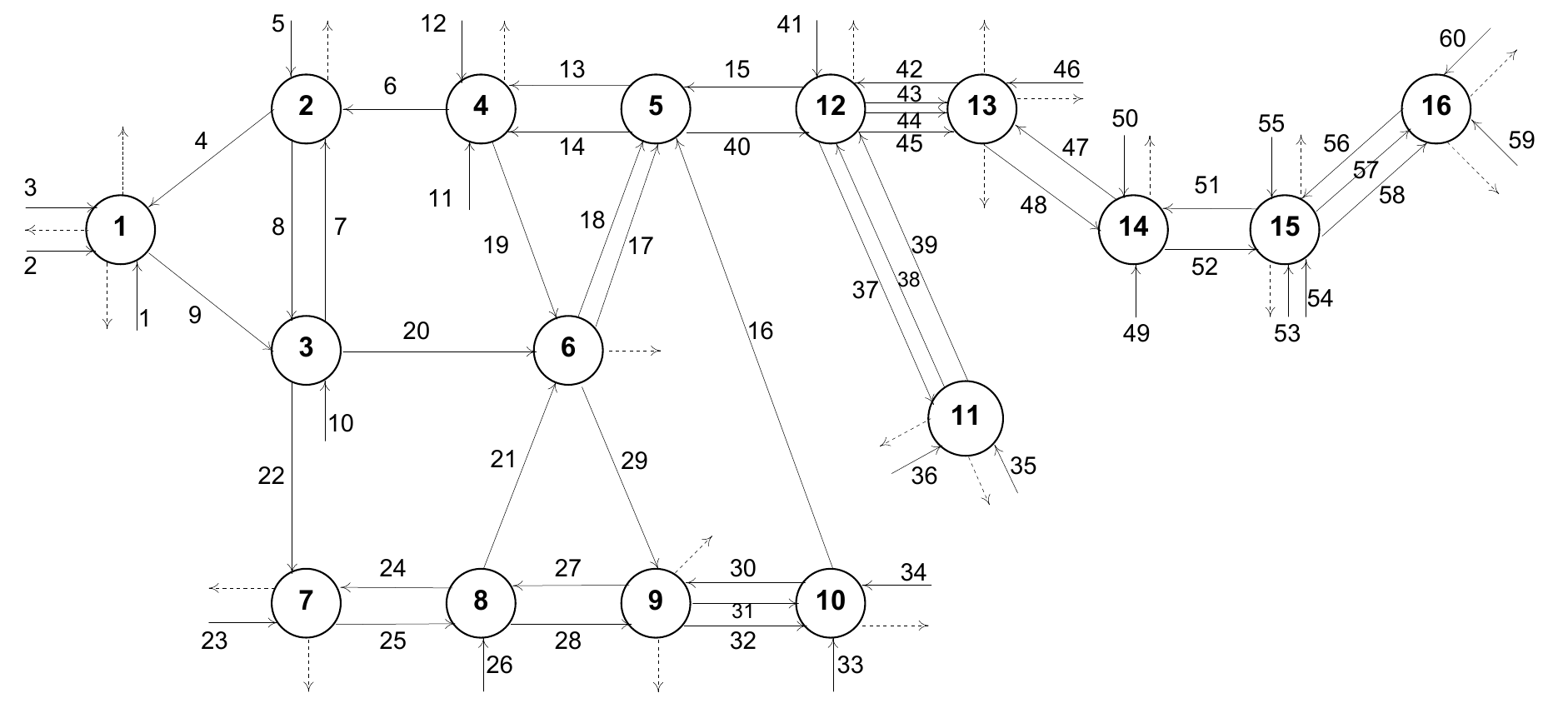}
	\caption{Chania urban traffic network topology graph \citep{PedrosoBatistaEtAl2022Saffron}.}
	\label{fig:Chania_60_42_Graph}
\end{figure}

To evaluate the performance of the proposed feedback-feedforward solution, it is numerically simulated for the urban traffic network of the city center of Chania, Greece,  whose topology graph is depicted in Fig. \ref{fig:Chania_60_42_Graph}. For that purpose, we make use of the network model and simulation tools made available in the SAFFRON toolbox \citep{PedrosoBatistaEtAl2022Saffron}. A MATLAB implementation of the proposed approach and all the source code of the simulations can be found as an example in the SAFFRON toolbox, available at {\small \href{https://github.com/decenter2021/SAFFRON/tree/master/Examples/PedrosoBatistaPapageorgiou2023}{\texttt{https://github.com/decenter2021/SAFFRON/tree/master/ Examples/PedrosoBatistaPapageorgiou2023}}}.

The nonlinear model \eqref{eq:LTI_sim_nl} is used to perform the macroscopic simulations. The Chania urban road network model comprises $J=16$ signalized junctions, and $L=60$ links. This network is feasible and a minimum complete stage strategy was used, whose details are omitted. The cycle time is $C=100$ s, the simulation sampling time is $T=5$ s, and the parameter that adjusts the sensibility of back-holding is $c_{\text{ug}}=0.85$. The sensing model used for simulation purposes is distinct from the one used for synthesis in \eqref{eq:sensor_output}. It is inspired by the sensor model employed in \citet[Section~IV.A]{vigos2010simplified} and accounts for the fact that the noise magnitude varies with the occupancy in the link and  it also takes into account noise that stems from the green-red signal switching pattern as colored noise, i.e.,
\begin{equation}\label{eq:sensor_output_sim}
	y_z^{\mathrm{sim}}(k_E) = x_z(k_E) + 0.05x_z(k_E) \psi(k_E) + 0.4 x_z(k_E) \phi(k_E),
\end{equation}
where $\psi$ is a unit zero-mean white Gaussian process and  $\phi$ is the band of a unit zero-mean white Gaussian process in the frequency range $[1/C;2/C]$.

We simulate four different methods to assess the performance of the proposed solution: (i)~ideal TUC, whereby the constant nominal historic demand is employed and perfect state feedback is applied, i.e., the controller is fed with the ground-truth link occupancy;  (ii)~ideal TUC-FF, whereby the feedback-feedforward controller proposed is employed resorting to perfect occupancy feedback  and perfect exogenous demand knowledge; (iii)~TUC-FF, whereby the feedback-feedforward controller is fed with the estimates from the proposed joint occupancy and demand estimation; and (iv)~TUC, whereby the constant nominal historic demand is employed and an estimation scheme analogous to the one proposed in this paper is used to compute occupancy estimates only. First, the ideal TUC and ideal TUC-FF methods are of paramount importance to evaluate the performance improvement that stems from the feedforward component of TUC-FF, regardless of the estimation performance. Second, TUC and TUC-FF are the methods that can be implemented in reality and offer a fair and realistic assessment of the magnitude of the overall performance improvement of TUC-FF versus lack thereof.

 The estimation sampling time is judiciously set to $E = 20\;\mathrm{s}$.  It is noted that, the higher the estimation sampling time, the higher the variance of the process noise in \eqref{eq:est_dyn}. Therefore, to reduce error propagation, it is desirable that $E$ is as low as possible. On the other hand, if $E$ is too low, the occupancy measurement error of the loop detectors will be very high, since there is only one loop detector in each road link. In \cite[Section IV.C]{vigos2010simplified}, the latter effect is studied in detail. Therein, it is concluded that a plateau of the occupancy measurement root mean square error (RMSE) is reached for estimation sampling times $E \geq 20\;\mathrm{s}$. Therefore, we choose the smallest $E$ that does not compromise the quality of the occupancy measurements, i.e., $E = 20\;\mathrm{s}$. The command action weighting matrix is set to $\mathbf{R} = 10^{-4}\mathbf{I}$, for both TUC and TUC-FF strategies. The noise covariance parameters employed to synthesize the estimator for the TUC-FF strategy were set, for every link $z\in \{1,\ldots,Z\}$, to ${R_z =(0.05x_{z,\text{max}}/4)^2}$, $Q_{x,z} = (S_zE/10)^2$, and $Q_{x,e} = (S_zE/10^3)^2$.  All these weighting parameters were tuned empirically by trial-and-error. The estimation solution implemented for the TUC strategy is a single-input-single-output Kalman filter. It filters the occupancy measurement modeled by \eqref{eq:sensor_output} using the occupancy dynamics
\begin{equation}
	x_z(k_E) = x_z(k_E) + (E/C)\mathbf{l_z}\!^\top\mathbf{B_u}\mathbf{u}_{\text{est},z}(k_E) + Ee^{\mathrm{\,hist}}_z + w_{\mathrm{x},z}(k_E), 
\end{equation}
whereby the process noise $w_{\mathrm{x},z}(k_E)$ is modeled by a zero-mean white Gaussian process whose covariance is also given by $Q_{\mathrm{x},z}$, and $e^{\,\mathrm{hist}}_z$ is the constant historical demand on link $z$.

\begin{figure}[t]
	\centering
	\includegraphics[width = \linewidth]{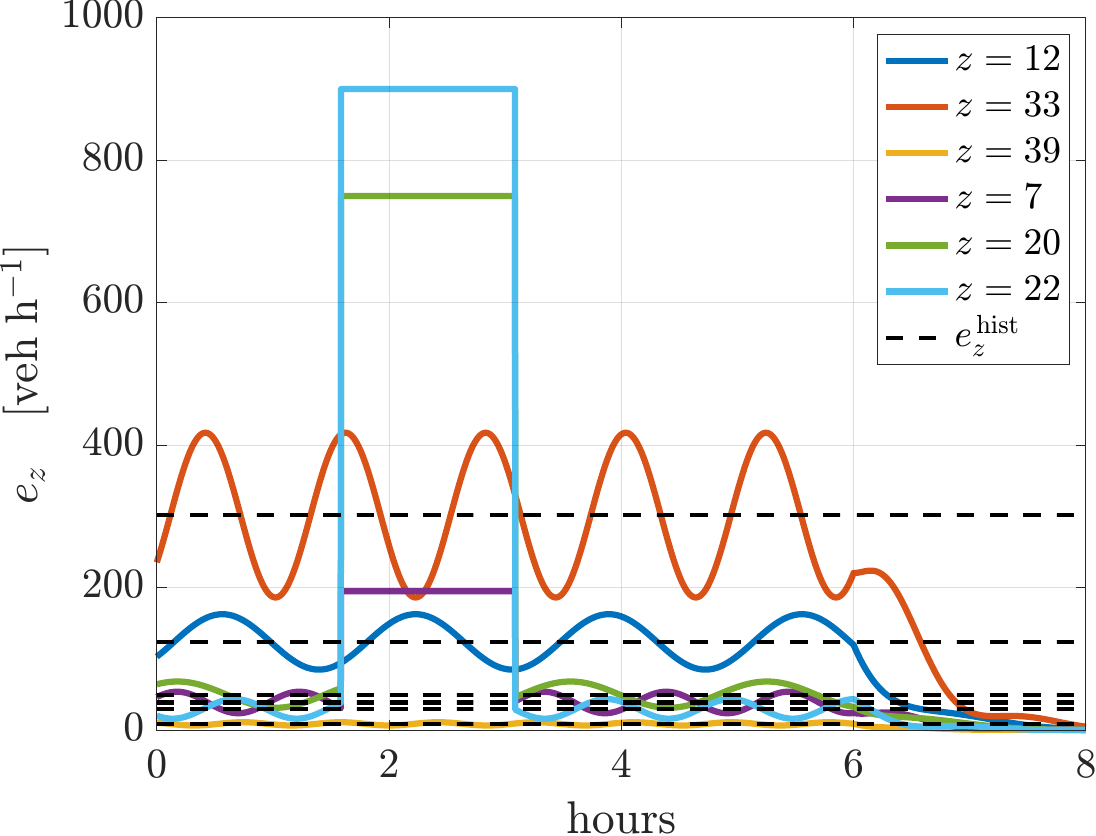}
	\caption{Exogenous demand in some links.}
	\label{fig:d_sc1}
\end{figure}

\begin{figure}[t]
	\centering
	\includegraphics[width = \linewidth]{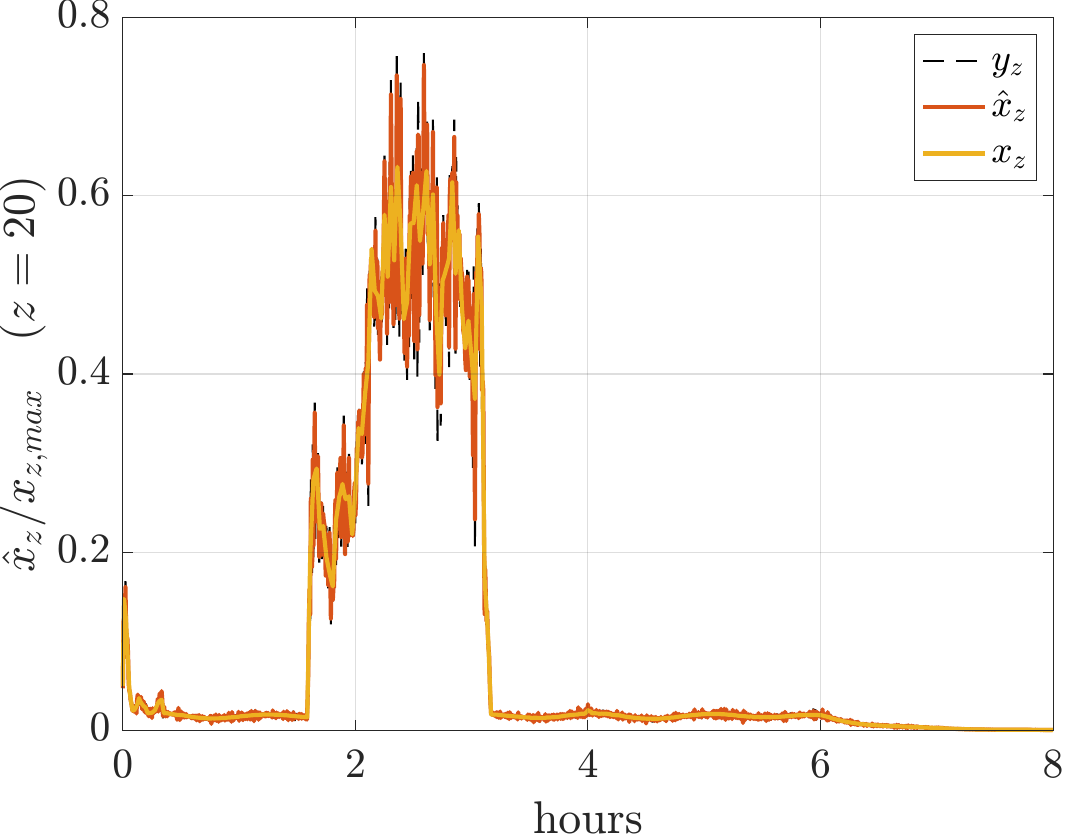}
	\caption{Comparison of estimate, ground-truth, and measurement of the occupancy of link $20$.}
	\label{fig:x_hat_sc1}
\end{figure}

\begin{figure}[t]
	\centering
	\includegraphics[width = \linewidth]{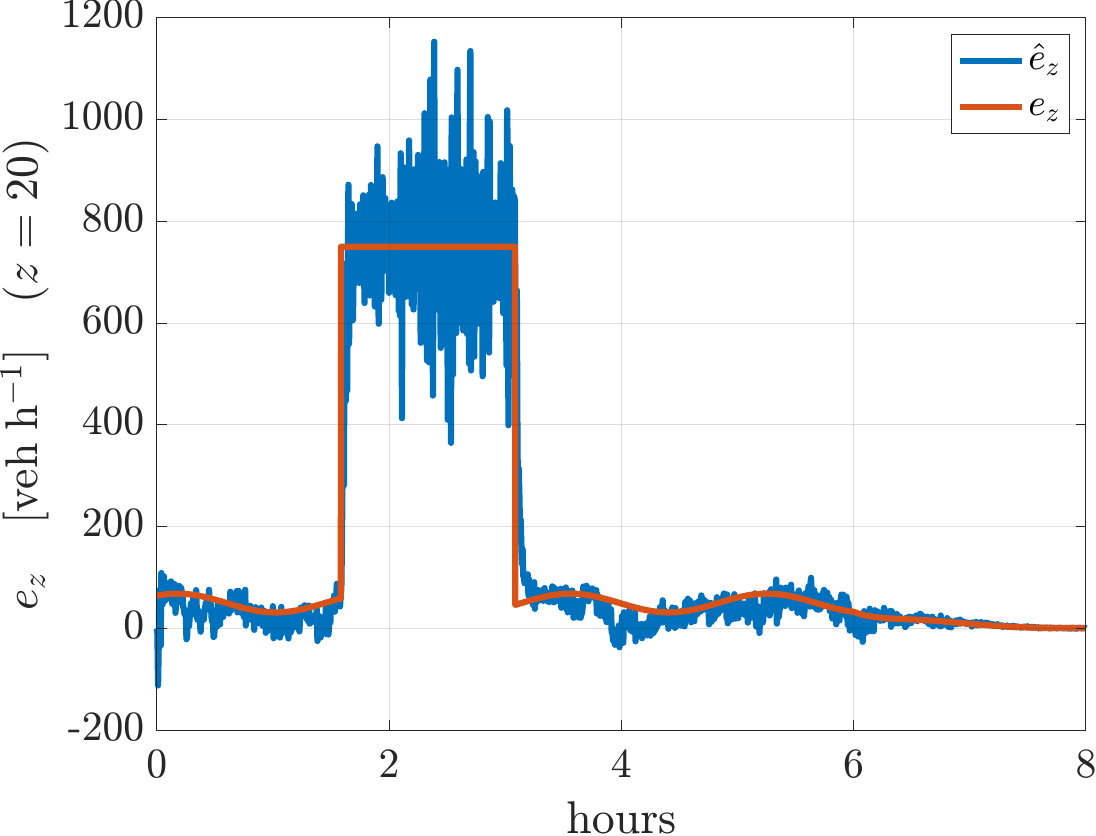}
	\caption{Evolution of the estimate of the exogenous demand on link $20$.}
	\label{fig:d_hat_sc1}
\end{figure}

In what follows, we present simulation results for two scenarios in the presence of dramatic demand patterns. First, we consider a simple synthetic scenario, whose analysis allows to gather insight into the proposed method. Second, we consider a more realistic scenario of a two-hour window on a weekday afternoon.

\subsection{Synthetic Exogenous Demand Scenario}%
In the first simulation scenario, the nominal historical demand $\{e^{\,\mathrm{hist}}_z\}_{z\in \{1,\ldots,Z\}}$ considered is constant. It is based on average real-world data and is available for the urban network of Chania in the SAFFRON toolbox \citep{PedrosoBatistaEtAl2022Saffron}. The ground-truth exogenous demand was set to a sinusoidal signal whose average is the nominal historical demand and whose amplitude, phase, and period were randomly generated from a uniform distribution between $[e^{\,\mathrm{hist}}_z/4; e^{\,\mathrm{hist}}_z/2]$, $[0;2\pi]$; and $[0.5;2]\;\mathrm{h}$, respectively. Moreover, to simulate the outflow of vehicles at the end of an event in the city center near junction $3$, the exogenous demand in the three links directed from junction $3$ (links $7$, $20$, and $22$), feature a pulse of high demand for $1.5\,\mathrm{h}$. In the last two hours of simulation, the exogenous demand is set to gradually fall to zero. An example of the evolution of the exogenous demand for some links is depicted in Fig.~\ref{fig:d_sc1}.

\begin{figure}[t]
	\centering
	\includegraphics[width = \linewidth]{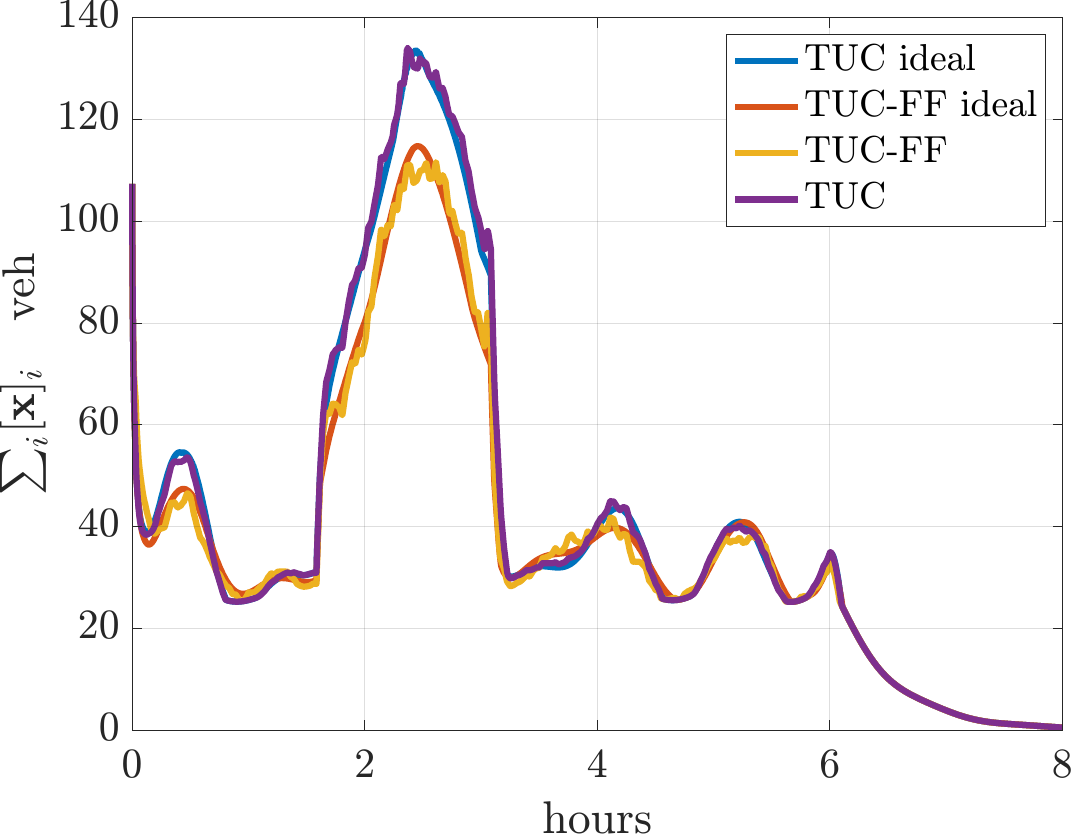}
	\caption{Evolution of total number of vehicles in the network.}
	\label{fig:sum_x_sc1}
\end{figure}

First, the evolution of the estimate, ground-truth, and measurement of the occupancy of link $20$ is depicted in Fig.~\ref{fig:x_hat_sc1}. The evolution of the estimate of the exogenous demand on link $20$ is depicted in Fig.~\ref{fig:d_hat_sc1}. One can readily point out that, although the occupancy estimate follows the ground-truth occupancy, the filtering performance is not very significant. That is due to the fact that the noise on the occupancy measurements simulated as described in \eqref{eq:sensor_output_sim} is proportional to the occupancy value, whereas the synthesis of the Kalman filter is based on a constant noise covariance. However, even though a gain scheduling technique could significantly improve estimation performance, it is evident in the sequel that the current estimation accuracy does not degrade the performance of the controller. The estimation of the exogenous demand also follows the ground-truth evolution.

Second, the evolution of the total number of vehicles in the network is depicted in Fig.~\ref{fig:sum_x_sc1} and the performance metrics are presented in Table~\ref{tab:metrics_sc1}. Comparing the performance of the ideal methods with the ones that make use of filtered estimates, one notices that the performance degradation is not significant. From Fig.~\ref{fig:sum_x_sc1}, one can immediately tell that during the periods of small variations of the exogenous demand w.r.t.\ the nominal value, the performance of the four methods is virtually identical. On the other hand, during the interval of the pulses in the exogenous demand in three links of the network, one can readily tell a significant performance improvement that stems from the feedforward component. Furthermore, the performance improvement on the RQB is striking, which is indicative that the feedforward component is indeed balancing the demand load significantly better than TUC.


\begin{figure}[t]
	\centering
	\includegraphics[width = \linewidth]{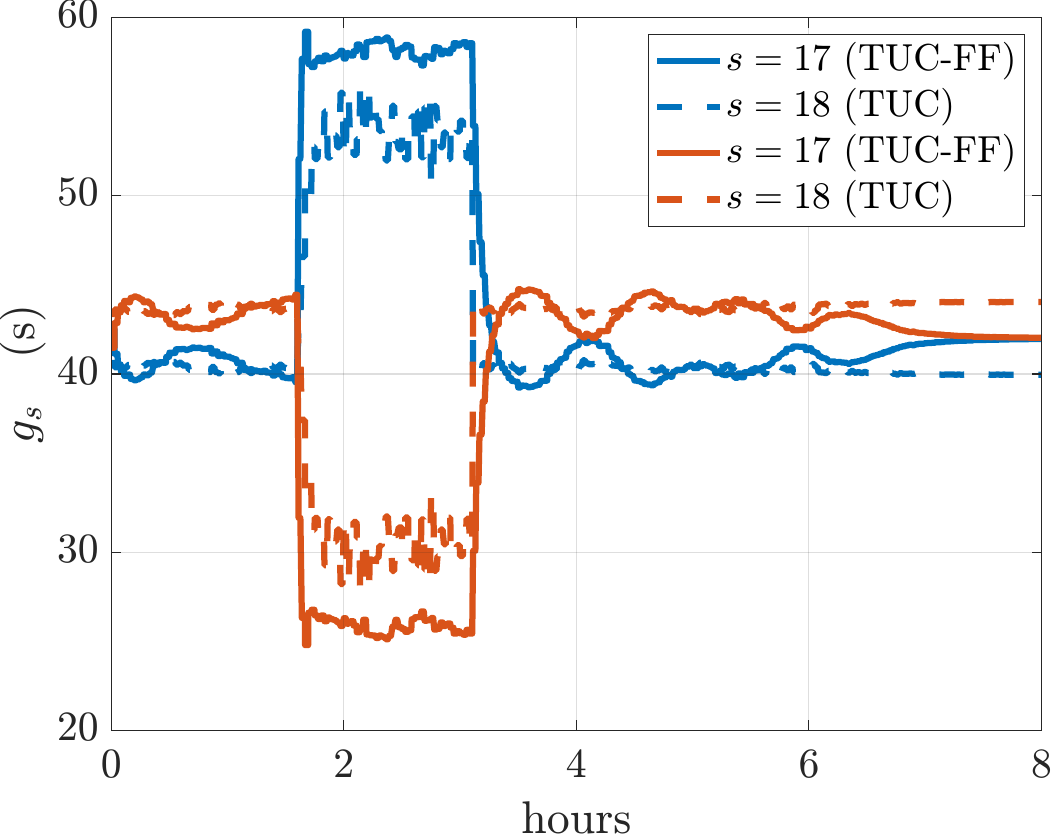}
	\caption{Evolution of green-times of the stages of junction $7$.}
	\label{fig:g7_sc1}
\end{figure}

\begin{table}[t]
	\centering
	\caption{Comparison of performance metrics for the first scenario.}
	\label{tab:metrics_sc1}
	\vspace{0.3cm}
	\renewcommand{\arraystretch}{1.2}
	\begin{tabular}{lcccc}
		\hline
		Method & TTS {$(\mathrm{veh}\cdot\mathrm{h})$}&  RBQ $\;\times 10^{-3}$ {$(\mathrm{veh})$}  & TTB\\
		\hline
		TUC ideal  & $360$ & $3.14$ & $0$ \\
		TUC-FF ideal & $307$ & $1.76$ & $0$ \\
		TUC-FF &  $306$ & $1.80$ & $0$ \\
		TUC & $365$ & $3.34$ & $0$ \\
		\hline
	\end{tabular}%
\end{table}

Third, Fig.~\ref{fig:g7_sc1} depicts the green-times of the stages $17$ and $18$, the stages of junction $7$, which is the junction downstream of link $22$. Stage $17$ gives r.o.w.\ to link $22$, and stage $20$ gives r.o.w.\ to links $23$ and $24$. Therefore, given a pulse of exogenous demand in link $22$, one would expect that the stage that gives r.o.w.\ to link $22$ is given more green time, which is exactly what happens, as seen in Fig.~\ref{fig:g7_sc1}.


\subsection{Realistic Exogenous Demand Scenario}

\begin{figure}[t]
	\centering
	\includegraphics[width = \linewidth]{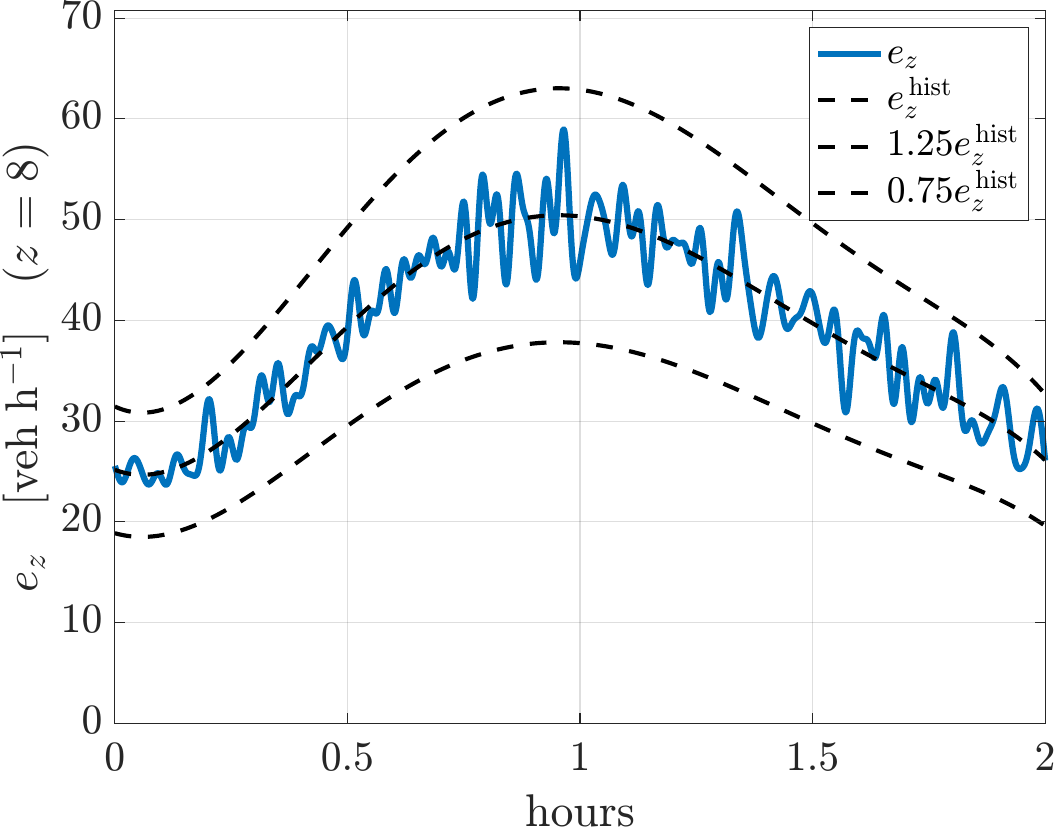}
	\caption{Realistic exogenous demand in link $8$ for a $2$ hour window (weekday afternoon).}
	\label{fig:d08_sc2}
\end{figure}

\begin{figure}[t]
	\centering
	\includegraphics[width = \linewidth]{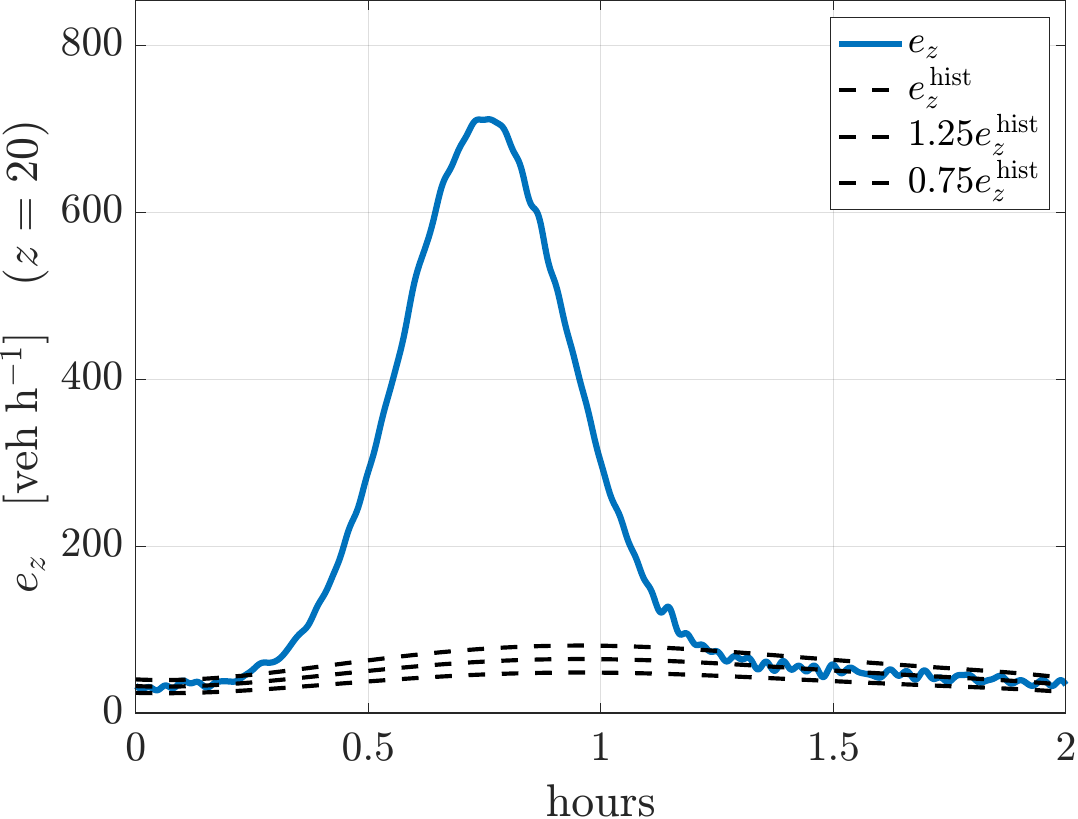}
	\caption{Realistic exogenous demand in link $20$ for a $2$ hour window (weekday afternoon), with a severe demand pulse.}
	\label{fig:d20_sc2}
\end{figure}

In the second simulation scenario, we consider a nominal historical demand $\{e^{\,\mathrm{hist}}_z\}_{z\in \{1,\ldots,Z\}}$ that follows a nominal profile of a weekday afternoon shown in \cite[Section II]{TettamantiLuspayEtAl2014}, whose average corresponds to the average real-world historical demands for Chania considered in the previous scenario. Similarly to \cite[Section II]{TettamantiLuspayEtAl2014}, the ground-truth exogenous demand corresponds to the nominal historical demand perturbed by multiplicative deviation. Similarly to the previous scenario, to simulate the outflow of vehicles at the end of an event in the city center near junction $3$, the exogenous demands in the three links directed from junction $3$ (links $7$, $20$, and $22$) feature a smooth pulse of higher demand. The evolution of the exogenous demand for links $8$ and $20$ is depicted in Figs.~\ref{fig:d08_sc2} and~\ref{fig:d20_sc2}, respectively.

\begin{figure}[t]
	\centering
	\includegraphics[width = \linewidth]{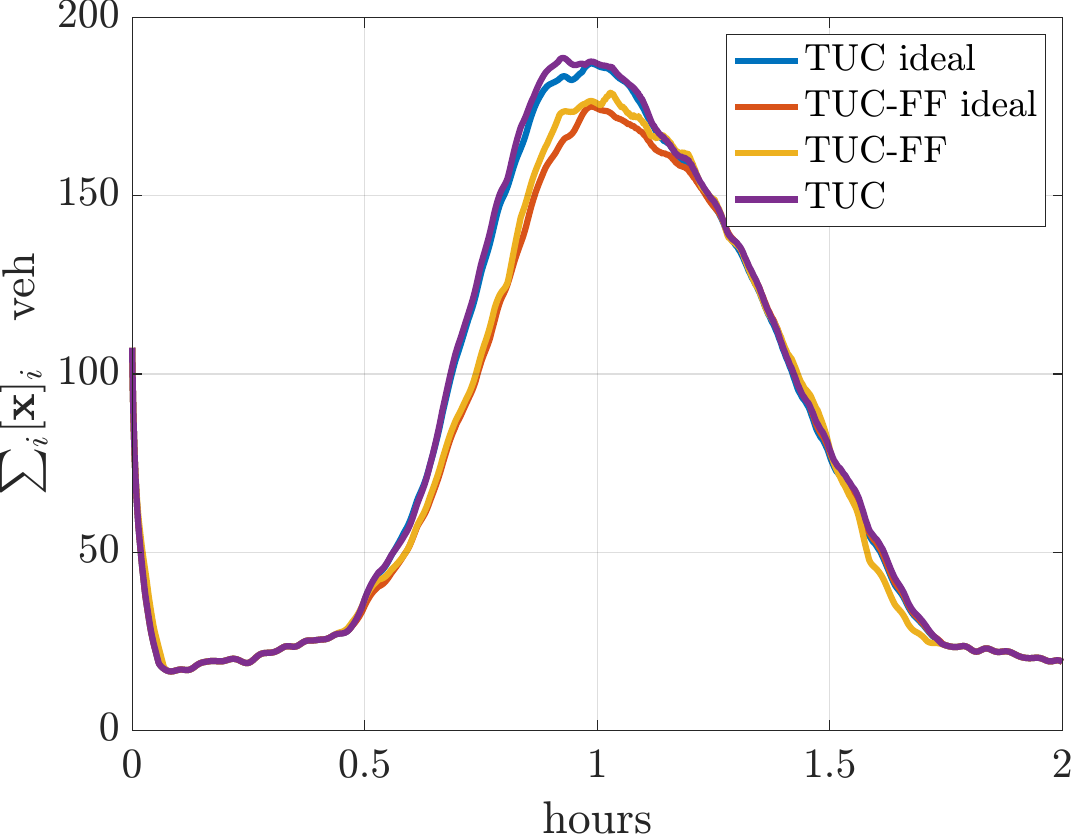}
	\caption{Evolution of total number of vehicles in the network.}
	\label{fig:sum_x_sc2}
\end{figure}

The evolution of the total number of vehicles in the network is depicted in Fig.~\ref{fig:sum_x_sc2} and the performance metrics are presented in Table~\ref{tab:metrics_sc2}. It is noticeable that the performance degradation due to the estimation algorithms is not significant.  Similarly to the previous scenario, during the periods of small variations of the exogenous demand w.r.t.\ the nominal value, the performance of the four methods is virtually identical. However, during the pulses in the exogenous demand, there is a significant performance improvement that stems from the feedforward component. Due to the smoothness of the pulse, the improvements in TTS and RQB are not as striking as in the first simulation scenario.

\begin{table}[t]
	\centering
	\caption{Comparison of performance metrics for the second scenario.}
	\label{tab:metrics_sc2}
	\vspace{0.3cm}
	\renewcommand{\arraystretch}{1.2}
	\begin{tabular}{lcccc}
		\hline
			Method & TTS {$(\mathrm{veh}\cdot\mathrm{h})$}&  RBQ $\;\times 10^{-3}$ {$(\mathrm{veh})$}  & TTB\\
		\hline
		TUC ideal  & $161$ & $2.24$ & $0$ \\
		TUC-FF ideal & $152$ & $2.08$ & $0$ \\
		TUC-FF &  $154$ & $2.01$ & $0$ \\
		TUC & $163$ & $2.25$ & $0$ \\
		\hline
	\end{tabular}%
\end{table}

\subsection{Discussion}

Overall, the simulation results show that, even though TUC has good disturbance rejection properties, during periods of high variation of the exogenous demand w.r.t.\ the nominal value, the proposed approach achieves significantly better performance. Three points are in order. First, it is important to stress that, although the methods in this paper are presented for a time-invariant triplet $(\mathcal{G},\mathbf{T}, \mathbf{t_0})$, the same approach can be followed to synthesize constant gains for distinct triplets that correspond to different nominal traffic patterns throughout the day. Second, the simulation results are presented for two exogenous demand patterns. The interested reader is encouraged to experiment with different exogenous demand patterns, which can be seamlessly accomplished with the source code that is made available. Third, although macroscopic simulations are widely accepted as a reliable assessment framework, further work should focus on more thorough simulations at the microscopic level.

%% file: sections/conclusion.tex
\section{Conclusion}\label{sec:conclusion}

In this paper, we model the traffic dynamics in a congested urban network with the store-and-forward model. First, we show that, with a single loop detector in each road link, it is possible to design an estimation approach to estimate both occupancy and net exogenous demand in every link. Second, using tools from optimal control theory, we synthesize a linear quadratic controller, whereby the exogenous demand is taken into account in the control law within a feedforward term. The resulting control scheme is a linear feedback-feedforward controller, which is fed with the occupancy and net exogenous demand estimates. Its simplicity allows for a seamless real-time implementation. Third, we assess the performance of the proposed estimation and control methods in the urban road network of Chania, Greece. By comparing our method with TUC, we conclude that for small perturbations about the historical nominal demand, there is no significant performance improvement, as expected. However, for realistic surges in the exogenous demand, one can observe a significant  improvement in performance as far as average travel time and relative balance of the network are concerned.

\section*{Acknowledgments}
This work was supported in part by LARSyS funding (DOI: {\small \href{https://doi.org/10.54499/LA/P/0083/2020}{\texttt{10.54499/LA/P/0083/2020}}, \href{https://doi.org/10.54499/UIDP/50009/2020}{\texttt{10.54499/UIDP/ 50009/2020}}}, and {\small \href{https://doi.org/10.54499/UIDB/50009/2020}{\texttt{10.54499/UIDB/50009/2020}}}), the Chinese NSF Project (52272334), the Ningbo International Science and Technology Cooperation Project (2023H020), and the National ``111" Centre on Safety and Intelligent Operation of Sea Bridges (D21013).

%% file: sections/appendix.tex
\appendix


\section{Proof of Lemma~\ref{lemma:fb-ff}}\label{app:proof_fb-ff}

The proof of this result follows the discrete-time procedures analogous to the ones described for continuous-time in \citet[Chapter~12.7]{papageorgiou2015optimierung} and \citet[Chapter~4]{lewis2012optimal}. For the sake of clarity of the procedure, we present the proof of the result incrementally using two stronger and more generic lemmas.

\begin{lemma}\label{lem:LTV_ff}
	Consider the LTV system 
	\begin{equation}
		\mathbf{x}_{k+1} = \mathbf{A}_k\mathbf{x}_k + \mathbf{B}_k\mathbf{u}_k + \mathbf{e}_k 
	\end{equation}
where $\mathbf{x}_k\in \mathbb{R}^n$ is the state, $\mathbf{u}_k\in \mathbb{R}^m$ is the input,  $\mathbf{e}_k \in \mathbb{R}^n$ is a time-varying known disturbance, and $\mathbf{A}_k$ and $\mathbf{B}_k$ are matrices of appropriate dimensions. Consider a finite-horizon cost function
\begin{equation}
	J_i = \frac{1}{2}\mathbf{x}_N^\top\mathbf{Q}_N\mathbf{x}_N + \frac{1}{2}\sum_{k = i}^{N-1} \left(\mathbf{x}_k^\top\mathbf{Q}_k\mathbf{x}_k+\mathbf{u}_k^\top\mathbf{R}_k\mathbf{u}_k\right),
\end{equation}
where $\mathbf{Q}_k \succeq \mathbf{0}$ and $\mathbf{R}_k \succ \mathbf{0}$ are matrices of appropriate dimensions and $\mathbf{x}_i$ is the fixed initial state. Then, the optimal control action $\mathbf{u}_k$ is given by
\begin{equation}\label{eq:LTV_opt}
	\mathbf{u}_k = -\mathbf{K}_k\mathbf{x}_k-\mathbf{K}_k^b(\mathbf{P}_{k+1}\mathbf{e}_{k}+\mathbf{b}_{k+1}),
\end{equation}
where
\begin{equation}\label{eq:rec_LTV}
	\begin{cases}
		\mathbf{P}_k = \mathbf{Q}_k + \mathbf{A}_k^\top\mathbf{P}_{k+1}(\mathbf{A}_k-\mathbf{B}_k\mathbf{K}_k), \quad \mathbf{P}_N = \mathbf{Q}_N\\
		\mathbf{b}_k = (\mathbf{A}_k-\mathbf{B}_k\mathbf{K}_k)^\top(\mathbf{b}_{k+1}+\mathbf{P}_{k+1}\mathbf{e}_{k}), \quad \mathbf{b}_N = \mathbf{0}\\
		\mathbf{K}_k = (\mathbf{R}_k+\mathbf{B}_k^\top\mathbf{P}_{k+1}\mathbf{B}_k)^{-1}\mathbf{B}_k^\top\mathbf{P}_{k+1}\mathbf{A}_k\\
		\mathbf{K}_k^b =   (\mathbf{R}_k+\mathbf{B}_k^\top\mathbf{P}_{k+1}\mathbf{B}_k)^{-1}\mathbf{B}_k^\top.
	\end{cases}
\end{equation}
\end{lemma}
\begin{proof}
	The final cost is given by 
	\begin{equation}
		\phi =  \frac{1}{2}\mathbf{x}_N^\top\mathbf{Q}_N\mathbf{x}_N
	\end{equation}
	and the Hamiltonian is given by
	\begin{equation}
		H_k = \frac{1}{2}(\mathbf{x}_k^\top\mathbf{Q}\mathbf{x}_k +\mathbf{u}_k^\top\mathbf{R}_k\mathbf{u}_k) + \boldsymbol{\lambda}_{k+1}^\top(\mathbf{A}_k\mathbf{x}_k+\mathbf{B}_k\mathbf{u}_k+\mathbf{e}_k),
	\end{equation} 
	where $\boldsymbol{\lambda}_{k+1}$ is the costate vector, which corresponds to a vector of Lagrange multipliers associated with the state evolution constraint. The final cost and the Hamiltonian can be used to obtain the state, costate, stationarity, and free final state optimality conditions which are, respectively, given by
	\begin{align}
		\mathbf{x}_{k+1} &= \frac{\partial H_k}{\partial \boldsymbol{\lambda}_{k+1}} =  \mathbf{A}_k\mathbf{x}_k + \mathbf{B}_k\mathbf{u}_k + \mathbf{e}_k,\\
		\boldsymbol{\lambda}_k &=  \frac{\partial H_k}{\partial \mathbf{x}_{k}} = \mathbf{Q}_k\mathbf{x}_k+\mathbf{A}_k^\top\boldsymbol{\lambda}_{k+1},\label{eq:costate}\\
		\mathbf{0} & = \frac{\partial H_k}{\partial \mathbf{u}_{k}} =  \mathbf{R}_k\mathbf{u}_k+\mathbf{B}_k^\top\boldsymbol{\lambda}_{k+1},
	\label{eq:stationarity} \\
		\boldsymbol{\lambda}_N &=  \frac{\partial \phi}{\partial \mathbf{u}_{k}} = \mathbf{Q}_N\mathbf{x}_N.\label{eq:free_final_state}
	\end{align}
For more details on how these conditions are obtained, refer to \citet[Chap.~4]{lewis2012optimal}. Now, we claim that one can write the costate as
\begin{equation}\label{eq:induction_hypothesis}
	\boldsymbol{\lambda}_k = \mathbf{P}_k\mathbf{x}_k + \mathbf{b}_k,
\end{equation}
for every $k\in \{i, \ldots, N\}$, where $\mathbf{P}_k$ is a symmetric positive definite matrix and $\mathbf{b}_k$ is a vector. We prove this claim by induction starting with $k = N$. From the free final state condition \eqref{eq:free_final_state}, one notices that $\boldsymbol{\lambda}_N$ can be written in the form of \eqref{eq:induction_hypothesis} whereby $\mathbf{P}_N = \mathbf{Q}_N$ and $\mathbf{b}_N = \mathbf{0}$. Now, consider a generic time instant $k$. From the stationarity condition in \eqref{eq:stationarity} and the induction hypothesis in \eqref{eq:induction_hypothesis} one has
\begin{equation}\label{eq:u_opt_proof}
	\begin{split}
		\mathbf{u}_k  &= - \mathbf{R}_k^{-1}\mathbf{B}_k^\top \boldsymbol{\lambda}_{k+1}\\
		& =  - \mathbf{R}_k^{-1}\mathbf{B}_k^\top ( \mathbf{P}_{k+1}\mathbf{x}_ {k+1}+ \mathbf{b}_{k+1})\\
		& =  - \mathbf{R}_k^{-1}\mathbf{B}_k^\top \mathbf{P}_{k+1}(\mathbf{A}_k\mathbf{x}_k+\mathbf{B}_k\mathbf{u}_k+\mathbf{e}_k) - \mathbf{R}_k^{-1}\mathbf{B}_k^\top\mathbf{b}_{k+1}\\
		& =  -\mathbf{K}_k\mathbf{x}_k-\mathbf{K}_k^b(\mathbf{P}_{k+1}\mathbf{e}_{k}+\mathbf{b}_{k+1}),
	\end{split}
\end{equation}
where 
\begin{equation}\label{eq:gains_proof}
	\begin{cases}
			\mathbf{K}_k = (\mathbf{R}_k+\mathbf{B}_k^\top\mathbf{P}_{k+1}\mathbf{B}_k)^{-1}\mathbf{B}_k^\top\mathbf{P}_{k+1}\mathbf{A}_k\\
		\mathbf{K}_k^b =   (\mathbf{R}_k+\mathbf{B}_k^\top\mathbf{P}_{k+1}\mathbf{B}_k)^{-1}\mathbf{B}_k^\top.
	\end{cases}
\end{equation}
From the costate stationarity condition in \eqref{eq:costate} and the induction hypothesis in \eqref{eq:induction_hypothesis} one has
\begin{equation}\label{eq:costate_expanded}
	\begin{split}
		\boldsymbol{\lambda}_k &=  \mathbf{Q}_k\mathbf{x}_k+\mathbf{A}_k^\top\boldsymbol{\lambda}_{k+1}\\
		\!\!\mathbf{P}_k\mathbf{x}_k + \mathbf{b}_k &= \mathbf{Q}_k\mathbf{x}_k+ \mathbf{A}_k^\top(\mathbf{P}_{k+1}\mathbf{x}_{k+1}+\mathbf{b}_{k+1})\\
		\!\!\mathbf{P}_k\mathbf{x}_k + \mathbf{b}_k &= 	 \mathbf{Q}_k\mathbf{x}_k+ \mathbf{A}_k^\top\mathbf{b}_{k+1}\\ +\mathbf{A}_k^\top\mathbf{P}_{k+1}&\left((\mathbf{A}_k-\mathbf{B}_k\mathbf{K}_k)\mathbf{x}_k+(\mathbf{I}-\mathbf{B}_k\mathbf{K}_k^b\mathbf{P}_{k+1})\mathbf{e}_k -\mathbf{B}_k\mathbf{K}_k^b\mathbf{b}_{k+1}\right)\!.\!\!\!\!\!\!\!\!\!\!\!\!\!\!
	\end{split}
\end{equation}
Since $\mathbf{x}_k$ is generally nonzero and since \eqref{eq:costate_expanded} is valid for all state sequences given $\mathbf{x}_i$, it follows that 
\begin{align}
	\!\!\!\mathbf{P}_k &= \mathbf{Q}_k + \mathbf{A}_k^\top\mathbf{P}_{k+1}(\mathbf{A}_k-\mathbf{B}_k\mathbf{K}_k)\label{eq:rec_P_proof}\\
	\!\!\!\mathbf{b}_k &= \mathbf{A}_k^\top \mathbf{b}_{k+1} + \mathbf{A}^\top_k\mathbf{P}_{k+1}\left( \mathbf{e}_k -\mathbf{B}_k\mathbf{K}_k^b\mathbf{b}_{k+1}  - \mathbf{B}_k\mathbf{K}_k^b\mathbf{P}_{k+1}\mathbf{e}_k\right)\!.\!\!\label{eq:rec_b_prev}
\end{align}
Furthermore, since $\mathbf{A}_k^\top -\mathbf{A}_k^\top\mathbf{P}_{k+1}\mathbf{B}_k\mathbf{K}_k^b = (\mathbf{A}_k-\mathbf{B}_k\mathbf{K}_k)^\top$, one may rewrite \eqref{eq:rec_b_prev} as
\begin{equation}\label{eq:rec_b_proof}
	\mathbf{b}_k =(\mathbf{A}_k-\mathbf{B}_k\mathbf{K}_k)^\top (\mathbf{b}_{k+1}+\mathbf{P}_{k+1}\mathbf{e}_k).
\end{equation}
Since $\mathbf{b}_k$ and $\mathbf{P}_{k}$ are well defined by \eqref{eq:rec_P_proof} and \eqref{eq:rec_b_proof}, the proof by induction is concluded. The result, thus, follows from the control action \eqref{eq:u_opt_proof} and \eqref{eq:gains_proof}, \eqref{eq:rec_P_proof}, and \eqref{eq:rec_b_proof}.
\end{proof}

\begin{lemma}\label{lem:LTI_ff}
	Consider the LTI system 
	\begin{equation}
		\mathbf{x}_{k+1} = \mathbf{A}\mathbf{x}_k + \mathbf{B}\mathbf{u}_k + \mathbf{e}
	\end{equation}
	where $\mathbf{x}_k\in \mathbb{R}^n$ is the state, $\mathbf{u}_k\in \mathbb{R}^m$ is the input,  $\mathbf{e}\in \mathbb{R}^n$ is a known constant disturbance, $\mathbf{A}$ and $\mathbf{B}$ are matrices of appropriate dimensions, and $\mathbf{x}_i$ is the fixed initial state. Consider an infinite-horizon cost function
	\begin{equation}
		J_\infty =  \frac{1}{2}\sum_{k = i}^{\infty} \left(\mathbf{x}_k^\top\mathbf{Q}\mathbf{x}_k+\mathbf{u}_k^\top\mathbf{R}\mathbf{u}_k\right),
	\end{equation}
	where $\mathbf{Q} \succeq \mathbf{0}$ and $\mathbf{R} \succ \mathbf{0}$ are matrices of appropriate dimensions. If the pair $(\mathbf{A},\mathbf{B})$ is controllable, then the optimal control action $\mathbf{u}_k$ is given by
	\begin{equation}
		\mathbf{u}_k = -\mathbf{K}\mathbf{x}_k-\mathbf{K}^e\mathbf{e},
	\end{equation}
	where $\mathbf{K}$ is given by the solution to the DARE
	\begin{equation}\label{eq:DARE_LTI}
		\begin{cases}
			\mathbf{P} = \mathbf{Q} + \mathbf{A}^\top\mathbf{P}(\mathbf{A}-\mathbf{B}\mathbf{K})\\
			\mathbf{K} = (\mathbf{R}+\mathbf{B}^\top\mathbf{P}\mathbf{B})^{-1}\mathbf{B}^\top\mathbf{P}\mathbf{A}\\
		\end{cases}
	\end{equation}
	and $\mathbf{K}^e$ is given by
	\begin{equation}
		\mathbf{K}^e = (\mathbf{R}+\mathbf{B}^\top\mathbf{P}\mathbf{B})^{-1}\mathbf{B}^\top(\mathbf{I}-(\mathbf{A}-\mathbf{B}\mathbf{K})^\top)^{-1}\mathbf{P}.
	\end{equation}
\end{lemma}
\begin{proof}
	This result is proved making use of Lemma~\ref{lem:LTV_ff}. Indeed, to particularize Lemma~\ref{lem:LTV_ff} to an infinite-horizon time-invariant setting, one may solve \eqref{eq:rec_LTV} asymptotically as $N \to \infty$. As a result, admitting $\mathbf{P} = \mathbf{P}_k = \mathbf{P}_{k+1}$ and $\mathbf{K} = \mathbf{K}_{k+1} = \mathbf{K}_{k}$ in  \eqref{eq:rec_LTV} yields the DARE \eqref{eq:DARE_LTI}, which is known to have a stabilizing solution since the pair $(\mathbf{A},\mathbf{B})$ is, by hypothesis, controllable. Likewise, admitting $\mathbf{b} = \mathbf{b}_k = \mathbf{b}_{k+1}$ in \eqref{eq:rec_LTV}, and taking into account that $\mathbf{A}-\mathbf{B}\mathbf{K}$ is Hurwitz and, thus, $(\mathbf{I}-(\mathbf{A}-\mathbf{B}\mathbf{K})^\top)$ is invertible, yields
	\begin{equation}
		\mathbf{b} = (\mathbf{I}-(\mathbf{A}-\mathbf{B}\mathbf{K})^\top)^{-1}(\mathbf{A}-\mathbf{B}\mathbf{K})^\top\mathbf{P}\mathbf{e}.
	\end{equation}
	Therefore, one may write the time-invariant analogous of \eqref{eq:LTV_opt} as
	\begin{equation*}
		\mathbf{u}_k = -\mathbf{K}\mathbf{x}_k-\mathbf{K}^e\mathbf{e},
	\end{equation*}
	where
	\begin{equation}
			\begin{split}
			\mathbf{K}^e & = (\mathbf{R}+\mathbf{B}^\top\mathbf{P}\mathbf{B})^{-1}\mathbf{B}^\top\left(\mathbf{I}+(\mathbf{I}-(\mathbf{A}-\mathbf{B}\mathbf{K})^\top)^{-1}(\mathbf{A}-\mathbf{B}\mathbf{K})^\top\right)\mathbf{P}\\
			& = (\mathbf{R}+\mathbf{B}^\top\mathbf{P}\mathbf{B})^{-1}\mathbf{B}^\top(\mathbf{I}-(\mathbf{A}-\mathbf{B}\mathbf{K})^\top)^{-1}\mathbf{P},
		\end{split}
	\end{equation}
	which concludes the proof.
\end{proof}

The proof of Lemma~\ref{lemma:fb-ff} follows immediately from the correspondence $\mathbf{x}_k <> \mathbf{x}(k)$, $\mathbf{u}_k <> \mathbf{g}(k)$, $\mathbf{A} <> \mathbf{I}$, $\mathbf{B} <> \mathbf{B_{g1}}$, $\mathbf{e} <> C\mathbf{\bar{e}_1}$, $\mathbf{K} <> \mathbf{K_1}$, and $\mathbf{K}^e <>\mathbf{K_{e1}}$ with Lemma~\ref{lem:LTI_ff}.  \hfill \ensuremath{\Box}